# Conformer-specific Chemistry Imaged in Real Space and Time

E. G. Champenois,[1,†] D. M. Sanchez,[1,2,†,‡] J. Yang,[1,3,4] J. P. F. Nunes,[5] A. Attar,[3] M. Centurion,[5] R. Forbes,[3] M. Gühr,[6] K. Hegazy,[1] F. Ji,[3] S. K. Saha,[5] Y. Liu,[7] M.-F. Lin,[3] D. Luo,[3] B. Moore,[5] X. Shen,[3] M. R. Ware,[1] X. J. Wang,[3,*] T. J. Martínez,[1,2,*] & T. J. A. Wolf[1,*]

**Affiliations:**
[1]Stanford PULSE Institute, SLAC National Accelerator Laboratory, Menlo Park, USA.
[2]Department of Chemistry, Stanford University, Stanford, USA.
[3]SLAC National Accelerator Laboratory, Menlo Park, USA.
[4]Center of Basic Molecular Science, Department of Chemistry, Tsinghua University, Beijing, China.
[5]Department of Physics and Astronomy, University of Nebraska-Lincoln, Lincoln, USA.
[6]Institut für Physik und Astronomie, Universität Potsdam, Potsdam, Germany.
[7]Department of Physics and Astronomy, Stony Brook University, Stony Brook, NY, USA.
*Corresponding authors. Emails: thomas.wolf@stanford.edu, toddjmartinez@gmail.com, wangxj@slac.stanford.edu
† These authors contributed equally to this work.

## Abstract:

Conformational isomers or conformers of molecules play a decisive role in chemistry and biology. However, experimental methods to investigate chemical reaction dynamics are typically not conformer-sensitive. Here, we report on a gas-phase megaelectronvolt ultrafast electron diffraction investigation of α-phellandrene undergoing an electrocyclic ring-opening reaction. We directly image the evolution of a specific set of α-phellandrene conformers into the product isomer predicted by the Woodward-Hoffmann rules in real space and time. Our experimental results are in quantitative agreement with nonadiabatic quantum molecular dynamics simulations, which provide unprecedented detail of how conformation influences time scale and quantum efficiency of photoinduced ring-opening reactions. Due to the prevalence of large numbers of conformers in organic chemistry, our findings impact our general understanding of reaction dynamics in chemistry and biology.

## Main Text:

Conformational isomers, or conformers, can interconvert via rotations around single chemical bonds. Interconversion between conformers represents an important step in many bi-molecular reactions, where reactant species must not only encounter one-another in a particular orientation, but also in a specific structural conformation.(*1*) Moreover, conformer dynamics arise naturally in self-ordering of macromolecular structures, e.g., protein folding.(*2*) However, investigation of conformer-specific dynamics on their natural time and length scales of femtoseconds and Ångströms is hindered by the insensitivity of established experimental methods to conformers.

The influence of conformers on photochemical reactivity is well-known, e.g. from electrocyclic reactions according to the Woodward-Hoffmann (WH) rules.(*3*) These rules predict them to be concerted and

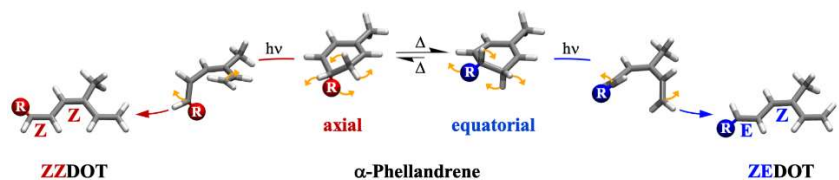

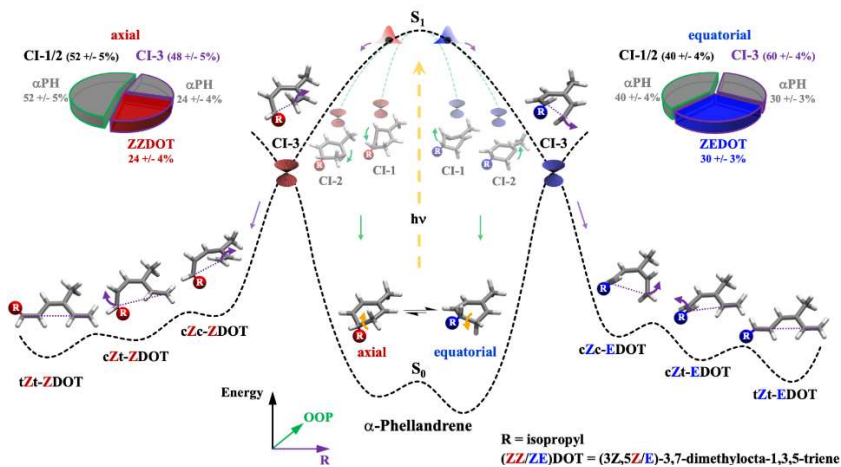

**Fig. 1. Conformer-Specific Photochemistry in a-Phellandrene.** a) Woodward-Hoffmann predictions for the conformer-specificity of photoinduced electrocyclic ring-opening in α-phellandrene. Its isopropyl substituent (R) can be in axial or equatorial orientation with respect to the carbon ring. Axial and equatorial conformers are in thermal equilibrium in solution phase (Δ).(*6*) The Woodward-Hoffmann rules predict a concerted, conrotatory ring-opening motion (orange arrows) yielding isomers with R in different positions depending on the reactant conformer. b) Schematic based on *ab initio* multiple spawning simulations of the photoinduced ring-opening. Equatorial and axial conformers are photoexcited from their respective ground state ($S_0$) energy minima to the first excited state ($S_1$). they evolve along an out-of-plane (green, OOP) towards conical intersections (CI) CI-1 and CI-2 or along the ring-opening coordinate (purple) towards CI-3. CI-1/CI-2 lead to reformation of a-Phellandrene while CI-3 leads to both αPH reformation and ring-opening. Several different conformers of the Z/EDOT photoproduct minima (cZc, cZt, and tZt) are accessible in the ground state. The two pie charts visualize the photoproduct distribution for axial and equatorial conformers as well as the distribution among the CI geometries CI-1 to CI-3 with errors representing 68% confidence intervals obtained from bootstrap analysis.

conformer-specific, i.e., leading to different reaction products depending on the reactant conformer. Electrocyclic reactions play an important role e.g., in chemical synthesis(*4*) and the Vitamin D production in human skin.(*5*)

An instructive example for conformer-specificity is the photochemical ring-opening of the monoterpene α-phellandrene (αPH),(*6*) which is produced by plants and used in the fragrance, food, and pharmaceutical industries.(*7*) α-Phellandrene consists of a 1,3-cyclohexadiene (CHD)-like ring moiety with two substituents, an isopropyl group at its $sp^3$ hybridized $C_1$ position and a methyl group at the $C_8$ position (see **Figs. 1** and **2**). The substitution gives rise to two conformers with the isopropyl group being in either quasi-axial (ax) or -equatorial (eq) orientation, (i.e., approximately perpendicular or parallel to the ring plane, respectively). The WH rules predict the ring-opening to take place in a concerted, conrotatory motion i.e., both ends of the newly formed open-ring molecule rotate away from each other in the same clockwise or counterclockwise direction (see **Fig. 1a**). This motion leads to different photoproducts depending on the reactant conformer, the isomers (3Z,5E)-3,7-dimethylocta-1,3,5-triene (ZEDOT, from eq-αPH) and (3Z,5Z)-dimethylocta-1,3,5-triene (ZZDOT, from ax-αPH, see **Fig. 1**).

Evidence for the conformer-specificity of the ring-opening in αPH(*6*) and other reactants(*8, 9*) has been observed in solution-phase measurements of photoproduct ratios. Moreover, resonance Raman investigations revealed evidence for the significance of conrotatory motion in the depopulation of the Franck-Condon region of the reactant excited state.(*10*) The photochemical dynamics leading from the Franck-Condon region to the ring-opened photoproducts of αPH have been investigated by several ultrafast spectroscopic studies in both the gas and condensed phase, however largely without sensitivity to the structural dynamics, reactant conformers, or photoproduct isomers.(*11-15*)

Here, we report on a combined megaelectronvolt ultrafast electron diffraction (UED) and *ab initio* multiple spawning (AIMS)(*16-18*) study of conformer-specific dynamics in the photochemical ring-opening of gas-phase αPH with unambiguous conformer-sensitivity confirming but going far beyond the qualitative orbital-symmetry based WH rule predictions. In both our static diffraction measurements and quantum chemical calculations, we find the eq-αPH conformer to dominate in our gas phase sample (see **supplementary note 1**). As opposed to all other studies of electrocyclic reactions, our time-resolved measurements directly image concerted and exclusive formation of the ZEDOT photoproduct predicted by the WH rules on the femtosecond timescale with sub-Å resolution and in quantitative agreement with AIMS simulations of the photochemical dynamics.

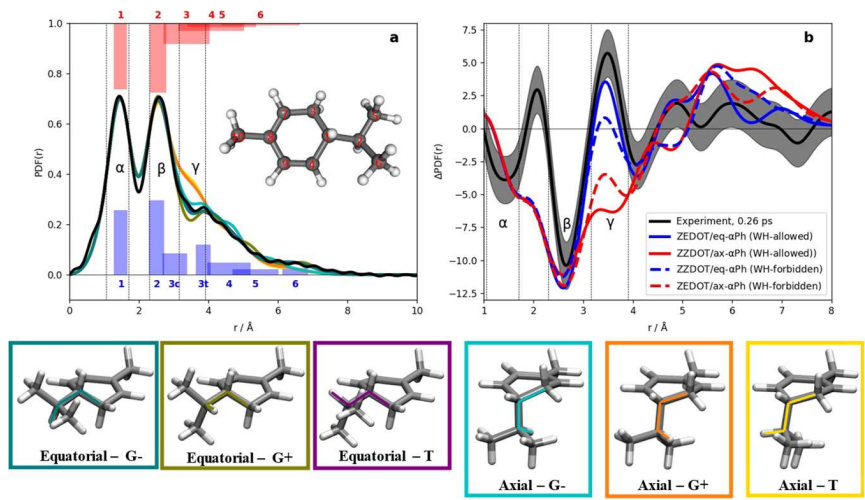

**Fig. 2. Comparison of experimental and simulated structural information.** a) Experimental (black) and simulated pair distribution functions PDF(r) of six α-phellandrene conformers, which are depicted below together with the dihedral angles defining the rotation of the isopropyl group. Carbon-Carbon coordination spheres for axial (red) and equatorial (blue) conformers are shown as bars. Additionally, the α, β, and γ ranges of **Fig. 3** are shown. The inset shows the Carbon atom numbering used in the text. b) Experimental difference PDF at a pump-probe delay of 0.26 ps (black) and simple simulations of the signature of Woodward-Hoffmann-(WH-)allowed and WH-forbidden reaction product signatures. Error bars represent a 68% confidence interval obtained from bootstrap analysis.

The simulations reveal that after photoexcitation into the lowest excited state ($S_1$) with $\pi\pi^*$ character, αPH non-adiabatically relaxes to the electronic ground state ($S_0$) through three different conical intersection (CI) geometries.(*19*) Only one of them yields ring-opened photoproducts while the other two form vibrationally "hot" reactant. The presence of the isopropyl group redirects significant amounts of the $S_1$ population towards the non-WH CIs, substantially reducing the WH-photoproduct yield compared to CHD.(*20*) Additionally, the rotational orientation of the isopropyl group gives rise to three rotational conformers (rotamers) for each of the eq and ax-αPH structures, labeled as gauche+/- (G+/G-) and trans (T, see **Fig. 2**), which are simultaneously present in our sample. These rotamers show significantly different ring-opening time scales.

**Figure 2a** shows the extracted static atomic pair distribution function (PDF(r)) from a gas phase sample of randomly oriented αPH molecules. For comparison, we plot simulated PDFs of all six ax-/eq-αPH rotamer geometries. We analyze the structural information contained in the PDFs in terms of Carbon coordination shells, which are additionally plotted as bars for the eq-αPH (blue) and ax-αPH (red) conformers in **Fig. 2a**:

The many C-C bonds (first Carbon coordination shell, labeled as α) in the system give rise to a strong maximum in all PDFs at 1.4 Å, while a second peak at 2.4 Å corresponds to the second Carbon coordination shell (the distances between Carbon atoms two atomic sites away from each other, β in **Fig. 2a**) with some contributions from the third coordination shell (distances of Carbons three atomic sites apart from each other). The signatures above 3 Å belong to the third and higher coordination shells.

We find the most significant differences between the PDFs within the third coordination shell (γ in **Fig. 2**). All eq-αPH and the ax-G-αPH rotamer as well as the experimental PDF exhibit a minimum around 3.5 Å, in contrast to the two remaining ax-αPH rotamers that produce a shoulder. The minimum results from a splitting of the third coordination shell in the eq-αPH and ax-G-αPH rotamers into distances between Carbons in cis-configuration (small dihedral angles, 3c in **Fig. 2a**) e.g., $C_3$-$C_9$, and trans-configuration (large dihedral angles, 3t in **Fig.2a**) e.g., $C_3$-$C_{10}$. The splitting collapses for the ax-G+ and ax-T rotamers due to one of the methyl groups of the isopropyl substituent being directly below the Carbon ring and thereby moving many of the ring-Carbon/methyl-carbon distances into the γ region (see structures in **Fig. 2**). The high level of agreement of the experimental PDF with those of eq-αPH and disagreement with some of the ax-αPH PDFs suggests the sample is dominated by eq-αPH rotamers in agreement with our quantum chemical calculations (see **supplementary note 1**) and previous results.(*21*)

The structural dynamics information contained in the time-dependent difference PDF (ΔPDF, difference between PDF(t) and steady-state PDF, see **Fig. 2b**) can be qualitatively understood from "simple" ΔPDF simulations generated from single photoproduct and reactant geometries. As shown in **Fig. 1b**, the ZEDOT and ZZDOT photoproducts can visit several ground state minima, which are all energetically accessible due to the high amount of nuclear kinetic energy in the photoproducts.(*22*) These minima differ by the dihedral angles between the double bonds, i.e., their orientation relative to each other with respect to the connecting single bonds. Each terminal double bond can be either in cis (c) or trans (t) orientation to the central double bond. It is conceivable that ZEDOT will be found more often in cZc-like configurations, since the geometry is closest to the eq-αPH reactant geometry. Moreover, like in the reactant geometry, the bulky isopropyl group is pointing away from the rest of the molecule. In ax-αPH, on the other hand, the conrotatory motion rotates the isopropyl group into the rest of the molecule, making cZc-like geometries significantly more unstable. The prevalence for cZc-ZEDOT (eq-αPH) and tZt-ZZDOT (ax-αPH) geometries is also reflected by our AIMS simulations (see below and **Fig. S1**). Therefore, the qualitative ΔPDF simulations are based on cZc-EDOT and tZt-ZDOT geometries.

**Figure 2b** shows an experimental ΔPDF 260 fs after photoexcitation with an ultrashort 266 nm optical pulse. We compare it with photoproduct ΔPDF simulations of the WH-predicted photoproducts eq-αPH→cZc-EDOT (blue line) and ax-αPH→tZt-ZDOT (red line). Additionally, we depict the WH-forbidden reactant/product ΔPDFs eq-αPH→tZt-ZDOT (blue dots) ax-αPH→cZc-EDOT (red dots). Atomic distance changes appear in ΔPDFs as a combination of a negative contribution at the initial atomic distance and a positive contribution at the delay-dependent distance. Accordingly, all ΔPDFs in **Fig. 2b** show signatures from ring-opening: The negative signature in the α-region directly follows from breaking the $C_1$-$C_3$ bond, while the negative signature in the β-region mainly results from increases in the $C_1$-$C_7$, $C_2$-$C_3$, and $C_3$-$C_4$ distances. Like the static PDFs, experimental and simulated ΔPDFs show the most prominent differences in the γ-region. Differences between experimental and simulated ΔPDFs in the α and β regions are due to the approximation of the photoproduct by a single geometry. The WH-allowed photoproduct of the eq-conformers shows good qualitative agreement with a strong positive signature in the γ-region of the experimental ΔPDF, whereas all other ΔPDFs show essentially zero or negative amplitude.

The difference between the simulated ΔPDFs can be explained in the same way as for the steady-state PDFs: The helical structure of the cZc-EDOT photoproduct results in a collapse of the distance splitting in the third coordination shell of the eq-αPH reactant leading to a positive ΔPDF signature in the γ region. In contrast, for the ax-αPH/tZt-ZDOT reactant/photoproduct combination, a splitting of the third coordination shell is induced by the more stretched photoproduct geometry leading to a negative ΔPDF signature. Both effects, the positive signature of eq-αPH/cZc-EDOT and the negative signature of ax-αPH/tZt-ZDOT, are direct results of the orientation of the isopropyl group in the reactant conformer and therefore an unambiguous signature of conformer-specific ring-opening.

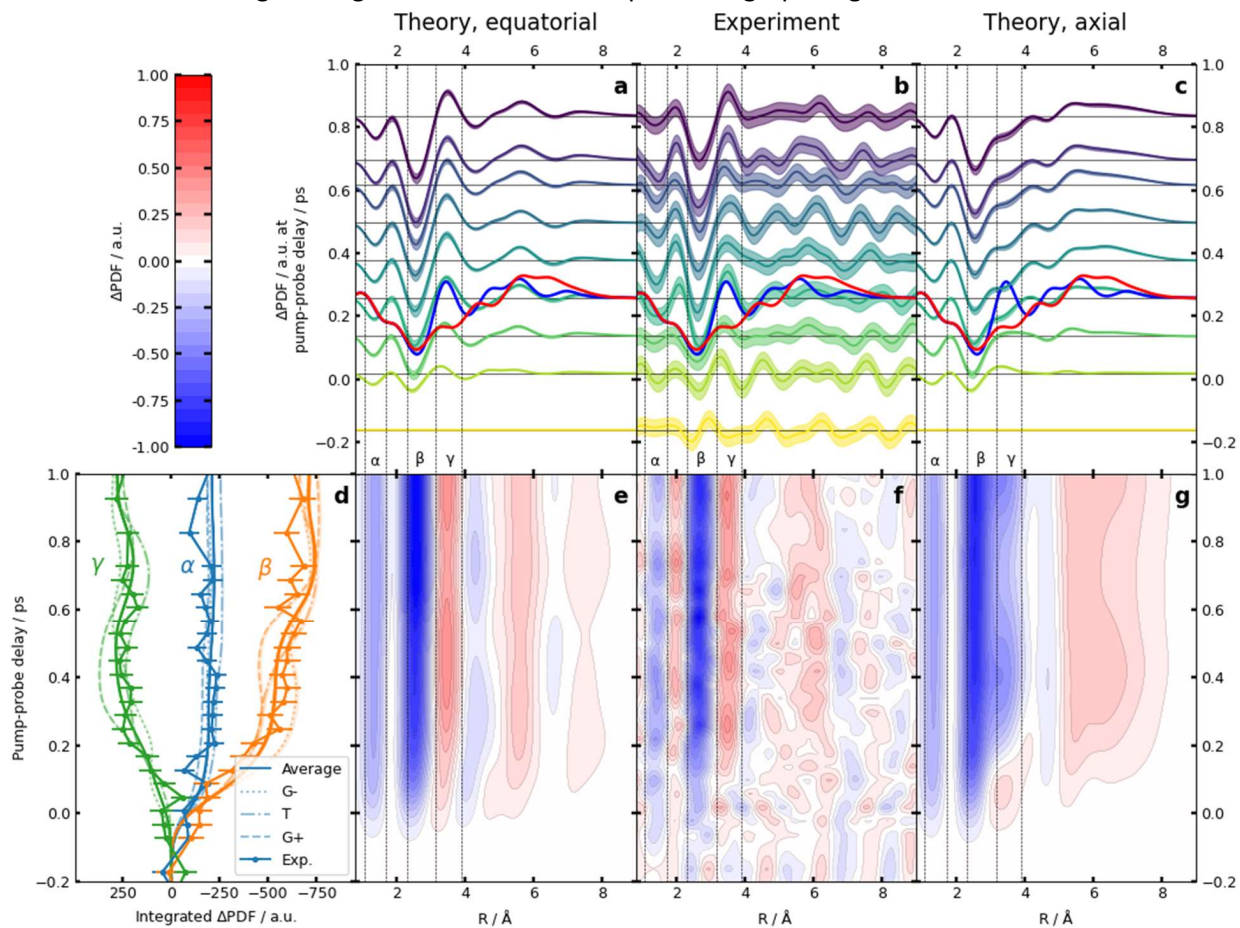

**Figure. 3: Comparison of experimental (b,d,f) and simulated time-dependent difference pair distribution functions (ΔPDF) of equatorial (a,d,e) and axial (c,g) conformers.** a,b,c) ΔPDFs at different pump-probe delays and e,f,g) false-color plots of ΔPDF over the whole investigated time window. Simulated ΔPDFs are based on ab-initio multiple spawning simulations (see methods). ΔPDFs at a delay of 260 fs in a,b,c are superimposed with the simulations of the conformer-specific Woodward-Hoffmann-allowed photoproducts for the axial (red) and equatorial (blue) conformers from **Fig. 2b**. d) Comparison of integrated signals in the α, β, and γ regions from the experiment, the averaged and the individual eq-rotamer simulations. Error bars represent a 68 % confidence interval obtained from bootstrap analysis. For the simulations, these error bars reflect convergence with respect to initial condition sampling.

The experimental ΔPDFs (**Fig. 3b** and **f**) are compared with those computed from AIMS simulations(*18*) (**Fig. 3a**, **c**, **e**, and **g**) using α-state-averaged complete active-space self-consistent field theory (α-CASSCF)(*23*) and assuming roughly equal contributions from the individual G-, T, and G+ rotamers at room temperature. The averaged AIMS ΔPDFs of the eq-conformers show quantitative agreement with the experimental data within their signal-to-noise level while the averaged AIMS ΔPDFs of the ax-conformers qualitatively disagree with the experimental data in the γ region. The experimental time-dependent signatures are clearly dominated by the eq-αPH rotamers (>80%, see **supplementary note 1**). Thus, we unambiguously observe exclusive formation of the WH-predicted photoproducts of the eq-αPH conformer within a few hundred fs after optical excitation. Our results constitute the first direct observation of conformer-specific chemical reaction dynamics in real space and time.

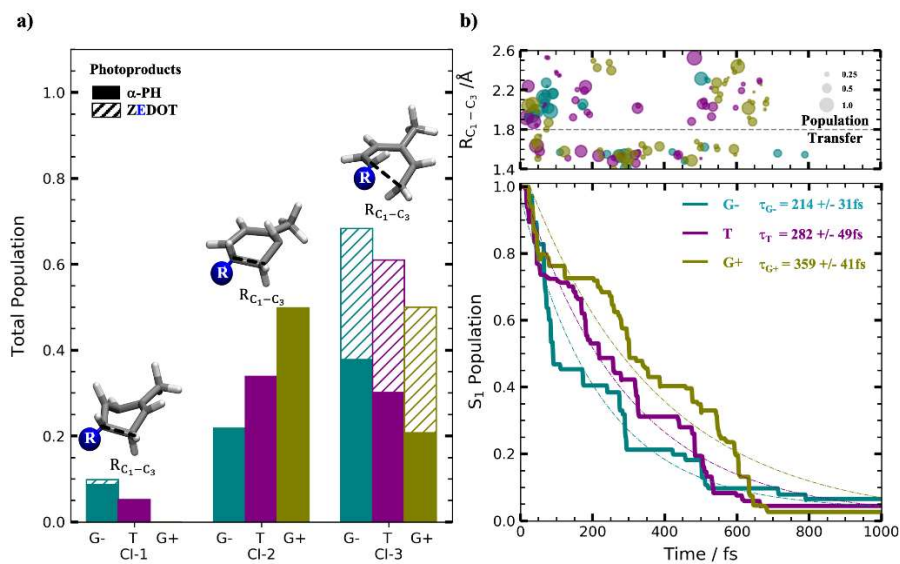

**Figure 4. Characterization of the Closed- and Open-Ring Nonradiative Relaxation Pathways of eq-Rotamers.** a) Histogram of the branching ratio between the closed and open pathways from simulations. Solid and striped bars represent fractional population that reform αPH or ZEDOT, respectively. b) (top) The $C_1$-$C_3$ distance vs population transfer time from $S_1$ to $S_0$ for all 45 eq initial conditions (see methods). The black dashed line corresponds to the threshold used to determine if the ring is open or closed. The circle radius is proportional to the amount of population involved in the transfer event and separated into the three rotamers. (bottom) The $S_1$ population decay for the three eq-rotamers in the first ps after photoexcitation.

Our AIMS simulations reveal additional aspects of the relaxation process, which cannot be easily extracted from the experimental data. α-Phellandrene relaxes via the WH-predicted conrotatory ring-opening motion through a CI (CI-3) yielding both ring-opened photoproducts and the closed-ring reactant (see **Figs. 1b, 4a**, and **S2**). Due to its differential nature, the ΔPDF observable is preferentially sensitive to the part of the population undergoing ring-opening. Additionally, out-of-plane bending in $S_1$ leads through two CIs (CI-1/CI-2) to the closed-ring reactant (**Table S1**). Furthermore, we observe a hitherto unknown secondary ring-closure reaction in the electronic ground state yielding 2-ladderane (5-isopropyl-2-methylbicyclo[2.2.0]hex-2-ene). The ring-closure is exclusively observed for the fraction of population undergoing internal conversion through CI-1 in the ground state (see **supplementary note 2**). Non-WH relaxation pathways have been observed before for the prototypical CHD molecule(*22*), but had marginal contributions to the relaxation process. The presence of the isopropyl substituent redirects 40±4 % of the eq-αPH $S_1$ population towards non-WH pathways and, thus, substantially reduces the ring-opening quantum yield (see **Fig. 1b**).

The time dependence of the measured and computed ΔPDFs in the α, β, and γ regions are shown in **Figs. 3d** and **S3b** for the eq and ax conformers, respectively. The averaged simulation assuming equal distribution of the three eq-rotamers (G-, T, and G+) leads to quantitative agreement with experiment, whereas individual eq-rotamer contributions to the ΔPDFs are quite distinct (see **Fig. 3d**). Thus, the rotational orientation of the isopropyl group must influence mechanism and time scale of ring-opening. This becomes obvious in **Fig. 4b**, top, where simulated non-adiabatic population transfer events are categorized with respect to their $C_1$-$C_3$ distances at the time of the transfer event i.e., their ability to undergo ring-opening and, thus, to significantly contribute to the ΔPDF observable (see above). In the case of the eq-G- rotamer, almost all population undergoing ring-opening relaxes to $S_0$ within 200 fs after photoexcitation. The other two rotamers show substantially less ring-opening in this timeframe. Their remaining population undergoes ring-opening significantly later, 500-700 fs after photoexcitation. The difference in ring-opening timescale is clearly reflected in the time-dependent ΔPDF amplitudes in **Fig. 3d.**

In summary, in the combination of UED with AIMS simulations, we directly observe conformer-specific reaction dynamics on their natural time and length scales in agreement with the Woodward-Hoffmann rules. Moreover, we observe considerable effects on the efficiency and timescale of the reaction from the presence and orientation of an isopropyl substituent of the molecule. Similar conformer-dependent photochemistry has been observed only in a limited number of studies of photoproduct distributions, mostly after conformer-specific resonance-enhanced photoionization.(*1, 24-26*) Our study marks to the best of our knowledge, the first time-domain observation of such conformer-specific photochemical dynamics. We speculate that the paucity of previous observations of conformer and rotamer-specificity is due to the absence of suitable experimental observables. These findings could have a large impact on our understanding of photochemical reaction mechanisms, since organic structures are prone to a large number of ground state conformers and rotamers. Future investigations of photochemical dynamics using single molecule methods with structural sensitivity, e.g. Coulomb explosion imaging at free electron lasers, might prove strongly conformer-specific photochemistry to be rather the rule than the exception.

## Acknowledgments:
We thank J. P. Cryan, M. C. Hoffmann, R. K. Li, M. Niebuhr, S. Weathersby, and T. Weinacht for their help and fruitful discussions. Lawrence Livermore National Laboratory is operated by Lawrence Livermore National Security, LLC, for the U.S. Department of Energy, National Nuclear Security Administration under Contract DE-AC52-07NA27344.

This work was supported by the AMOS program within the US Department of Energy, Office of Science, Basic Energy Sciences, Chemical Sciences, Geosciences, and Biosciences Division.

The experimental part of this research was performed at the SLAC MeV UED facility, which is supported in part by the DOE BES SUF Division Accelerator & Detector R&D program, the Linac Coherent Light Source (LCLS) Facility, and SLAC under contract nos. DE-AC02-05-CH11231 and DE-AC02-76SF00515.

M.G. is funded via a Lichtenberg Professorship of the Volkswagen Foundation.

D.M.S. is grateful to the NSF for a graduate fellowship.

M.C. is supported by the US Department of Energy Office of Science, Basic Energy Sciences under award no. DE-SC0014170.

Y.L. is supported by the US Department of Energy Office of Science, Basic Energy Sciences under award no. DE-FG02-08ER15984

# Supplementary Information

Conformer-specific Chemistry Imaged in Real Space and Time

E. G. Champenois,[1,†] D. M. Sanchez,[1,2,†,‡] J. Yang,[1,3,4] J. P. F. Nunes,[5] A. Attar,[3] M. Centurion,[5] R. Forbes,[3] M. Gühr,[6] K. Hegazy,[1] F. Ji,[3] S. K. Saha,[5] Y. Liu,[7] M.-F. Lin,[3] D. Luo,[3] B. Moore,[5] X. Shen,[3] M. R. Ware,[1] X. J. Wang,[3,*] T. J. Martínez,[1,2,*] & T. J. A. Wolf[1,*]

## Methods

### Ultrafast electron diffraction experiments:

The experimental apparatus is described in detail elsewhere.(*27*) In short, we use the 800 nm output of a Ti:Sapphire laser system and separate two beam paths. Pulses in both beam paths are frequency-tripled. The pulses of the probe beam path are directed onto the photocathode of an RF gun and eject an ultrashort pulse containing ~$10^4$ electrons. 3.7 MeV electrons are generated using a S-band photocathode radio frequency (RF) gun(*28*) and focused through a holey mirror to a spot size of 200 µm FWHM in the interaction region of a gas phase experimental chamber. The pump pulses (7 µJ) are focused into the experimental chamber to a diameter of 240 µm FWHM and overlapped with the electron pulses at a 2° angle. The experimental response function including effects of the optical and electron pulse length as well as relative arrival time jitter is estimated to be 150 fs.(*22*) αPH is purchased from Sigma-Aldrich and used without further purification. We use a static-filled 3 mm flow cell (550 µm orifices, sample at room temperature) in combination with a repetition rate of 360 Hz. Diffracted electrons are detected by a combination of a phosphor screen and an EMCCD camera. Based on the relative static and dynamic signal levels, we estimate to excite about 2.5 % of the molecules (see **Fig. S4**). Time-dependent diffraction is measured at a series of time delay points between -2 ps and +1 ps in each scan. The separation between time delay points is 50 fs, except for the earliest and latest delay points, where it was considerably larger. At each time delay point, we integrate diffraction signal for 10 seconds. The full data set includes 88 such scans. The sequence of delay steps is randomized for every scan to avoid systematic errors.

Generation of modified molecular diffraction and pair distribution functions from experimental data:

Determination of modified molecular diffraction (sM(s), see **Fig. S5**) from 2-dimensional molecular diffraction data is described in detail in the supplementary information of Ref. (*22*). Similar to Ref. (*22*), generation of static atomic pair distribution functions (PDFs, see **Fig. 2a**) requires extrapolating the experimental sM(s) for s≤0.8 Å$^{-1}$ with simulated sM(s) signal (see **Fig. S5**). As evident from **Fig. S5**, the extrapolated range (dashed black line) does not show conformer-dependence in the simulations. Thus, our mode of extrapolation does not bias the shape of the PDF towards a specific conformer.

ΔsM(s, t) is generated by subtracting sM(s) before time zero (before the onset of transient features) from sM(s) of all pump-probe delays. As for PDFs, extrapolation of these traces to s=0 Å$^{-1}$ is required to avoid artifacts in ΔPDF(r,t). In this low s region, the traces are set to ΔsMhole(s,t)= ΔsMeq(s,t=1ps)×(1 + erf[(t-t0)/τ]). The first term is the average of the simulated ΔsM(s,t) traces for the three equatorial conformers at a delay of 1 ps. The second term sets the time-dependence, which is assumed to follow a simple error function with the onset time t0 and width τ found by curve-fitting to ΔsMeq(s,t). As described in detail in the supplementary material of Ref.(*22*), for the generation of ΔPDF(r,t), the high s contributions of

ΔsMeq(s,t) are smoothly damped using a Gaussian function $e^{-ks^2}$ with k=0.028 Å$^2$. To ensure that the low s extrapolation of the experimental data does not bias the ΔPDF(r,t), we generate ΔPDF(r,t) for both the simulated averaged eq conformers and the experimental data, where we set the low s range to zero (see **Fig. S6**).

### Theoretical Method:

AIMS simulations interfaced with GPU-accelerated α-CASSCF(*23, 29-31*) are used to model the photodynamics of isolated ax and eq rotamers of ⍺PH. The α-CASSCF has shown to be well-suited for this system based on our previous CHD work and single-point XMSPT2 calculations (**Fig. S7-9**).(*22*) Our active space consists of six electrons in four orbitals determined to minimize the average energy of the lowest two singlet states, within the 6-31G* basis set, i.e. α-SA-2-CASSCF(6,4)/6-31G*. Electronic structure calculations are performed with TeraChem.(*32-34*) Following previous work, we use an α value of 0.82. A total of 90 initial conditions (15 sets of positions and momenta for each conformer) were selected from the computed electronic absorption spectrum (**Fig. S10**) and used to initiate the AIMS dynamics. The active-space molecular orbitals (MO) for all isomers were nearly identical to CHD (**Fig. S11**). These ICs are then placed on the S$_1$ surface and propagated with AIMS.

The first two singlet states (S$_0$ and S$_1$) are included in the dynamics. All required electronic structure quantities (energies, gradients, and nonadiabatic couplings) are calculated as needed with α-SA-2-CASSCF(6,4)/6-31G*. An adaptive timestep of 0.48 fs (20 au) (reduced to 0.12 fs (5 au) in regions with large nonadiabatic coupling) is used to propagate the centers of the trajectory basis functions (TBFs). A coupling threshold of 0.01 au (scalar product of nonadiabatic coupling and velocity vectors) demarcates spawning events generating new TBFs on different electronic states. Population transfer between TBFs is described by solving the time-dependent Schrödinger equation in the time-evolving TBF basis set.

We simulate the ultrafast dynamics for the first 1 ps of all six equatorial and axial conformational isomers of ⍺-PH by: 1) using AIMS to propagate the initial wavepacket for the first 500 fs or until all population has returned to the ground state, 2) stopping TBFs on the ground state when they are decoupled from other TBFs (off-diagonal elements of the Hamiltonian become small), and 3) adiabatically continuing these stopped TBFs using the positions and momenta from the last frame in AIMS as initial conditions for adiabatic molecular dynamics with unrestricted DFT using the Perdew-Burke-Ernzerof hybrid exchange-correlation functional,(*35*) i.e uPBE0/6-31G*. A total of 398 TBFs are propagated, with 306 of these being adiabatically continued on the ground state with DFT.

### Simulation of modified molecular diffraction and pair distribution functions:

The sM(s) simulations within the independent atom model (IAM) are generated from molecular geometries using a publicly available python code(*36*) and atomic scattering functions from the elsepa program.(*37*) PDFs are generated from the simulated sM(s) using the same code as for the experimental data.

For the creation of ΔPDFs from the AIMS simulations, sM(s) functions are evaluated for each time step of the simulation and each trajectory basis function (TBF) of a given initial condition separately both for the portion of the simulation using α-CASSCF and the extension on the ground state surface with DFT. The sM(s) functions of different TBFs are averaged for each time step according to their population weights. The resulting averaged time-dependent sM(s) functions are rebinned to 2 fs time steps. ΔsM(s, t) of a specific conformer are created by averaging the sM(s, t) from all initial conditions of this conformer and subtracting the initial sM(s, t=0) function from the average. The ΔPDF(r, t) functions are created from ΔsM(s, t) functions using the same code as for the experimental ΔPDFs. To account for the experimental response function, the ΔPDFs are convolved with a 150 fs FWHM Gaussian in time.

# Supplementary Text

## Supplementary note 1

We investigate the ground state potential energy surface using a number of different methods. A summary of our results can be found in **Fig. S12**. Our quantum chemical calculations find six possible αPH minimum geometries, three different rotamers for eq and ax geometries, which we label according to the $C_3$-$C_1$-$C_2$-$H_{ISO}$ dihedral angle of the isopropyl group shown in **Fig. 2** (i.e. gauche- (G-), trans (T), and gauche+ (G+)). Their $S_0$ minima are at -58.5°/-70.7° (G-), -172.9°/179.4° (T), and 57.0°/48.9° (G+) for ax/eq conformers, respectively. Our calculations find the eq-αPH minimum geometries to be in general more stable than ax-αPH and the thermal equilibrium, thus, dominated by similar fractions of the three eq-αPH rotamers. This is confirmed by a recent study on matrix isolated αPH which could only identify vibrational signatures of equatorial conformers.(*21*)

To quantify the possible amount of axial conformers in our gas phase sample, we fit a linear combination of the simulated sM(s) functions of all six conformers to the experimental sM(s). The coefficients of the axial conformer sM(s) functions are always negative or zero if forced to be positive semidefinite. Thus, we cannot find any direct evidence for the presence of axial conformers in the static diffraction. We estimate the uncertainty of this assessment to be on the order of 10 %,

To quantify the contribution of ax-αPH ring-opening to the experimental signal, we create linear combinations of the averaged eq-αPH and ax-αPH and compare the time-dependence of the integrated α, β, and γ regions to the experimental data, analogous to **Fig. 3d**. A comparison using a linear combination with 20% ax-αPH contribution is plotted in **Fig. S13**. The α and β regions are only moderately sensitive to the fraction of ax-αPH contribution, since both ax- αPH and eq-αPH open the ring. As expected from **Fig. S3**, the γ-region is significantly more sensitive to the fraction of ax-αPH contribution. In **Fig. S13**, the intensity of the γ-signature is reduced far enough to be outside the error bars (68% confidence interval) of the experiment. Therefore, we estimate the contribution of ax-αPH to the time-dependent experimental signal to be <20%.

## Supplementary note 2

In the AIMS simulations, we observe an additional photoproduct, 2-ladderane (5-isopropyl-2-methylbicyclo[2.2.0]hex-2-ene), which is formed in the electronic ground state exclusively by population having undergone internal conversion through the conical intersection CI-1. Since we do not observe internal conversion through CI-1 for the eq-G+ and ax-G- rotamers, the product is exclusively observed for the eq-T, eq-G-, ax-T, and ax-G+ rotamers (see **supplementary movies 1-4**). The ΔPDF signatures of 2-ladderane and ZEDOT are quite similar (see **Fig. S14**), especially in the γ-region. Additionally, the simulations predict only 5 % of the eq-α-phellandrene population undergoing internal conversion through CI-1 (see **Fig. 4** and **Table S1**). Therefore, a direct signature from the 2-ladderane contribution would be difficult to extract from the experimental data. However, it is included in the simulations and, therefore contributes to the quantitative agreement with experiment.

# Supplementary Figures

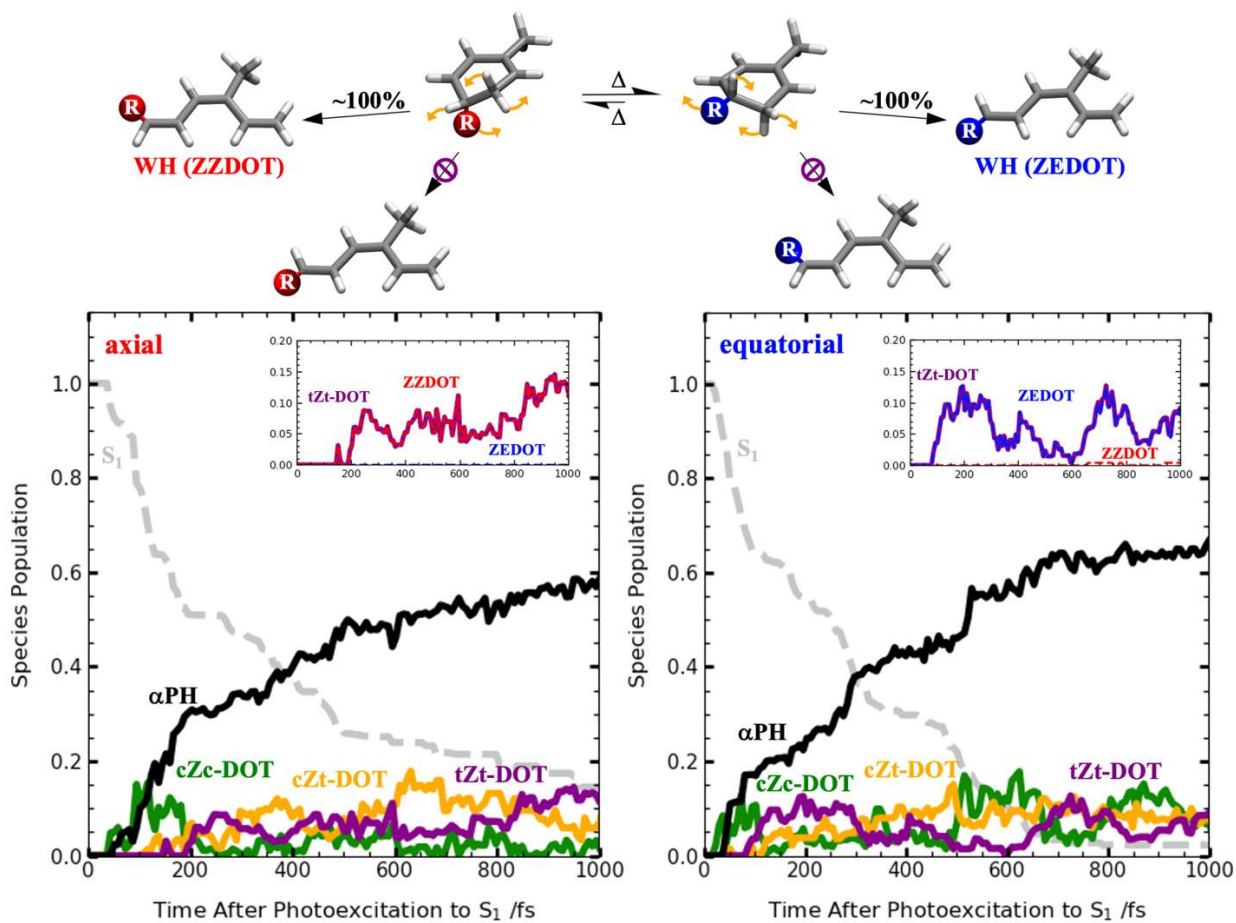

**Fig. S1.** AIMS Wavepacket Population for Axial and Equatorial Conformers. (Top) The Woodward-Hoffman allowed photoproducts from ax and eq aPH (the isopropyl group is represented as the R group). (Bottom) The wavepacket population from the axial (left) and equatorial (right) AIMS dynamics for the first picosecond after photoexcitation for all 90 ICs (45 each for ax and eq) considered. Snapshots every 5fs were binned based on the aPH, cZc, cZt, and tZt configurations and weighted according to their amplitudes. The insets show the tZt-DOT population decomposed into ZZDOT and ZEDOT contributions for ax (left) and eq (right) photoproducts. Conrotatory ring-opening in the ax and eq ICs leads almost exclusively to the WH predicted ZZDOT (red) and ZEDOT (blue) photoproducts, respectively.

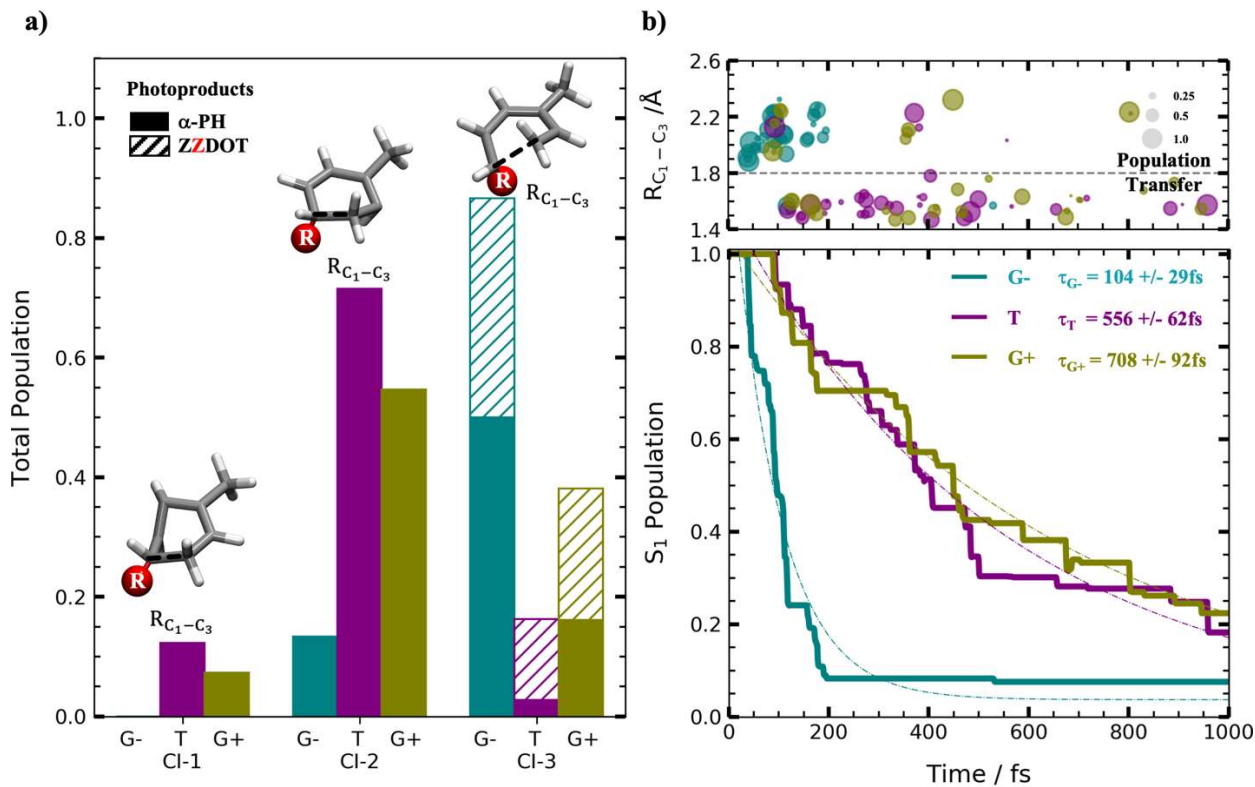

**Fig. S2.** Characterization of the Closed- and Open-Ring Nonradiative Relaxation Pathways for ax-Rotamers. a) A histogram of the branching ratio between the closed and open pathways from AIMS dynamics. Solid and striped bars represent fractional population that reformed αPH or cZc-ZDOT, respectively. b) (top) The $C_1$-$C_3$ distance vs spawning time for all 45 ax initial conditions. The black dashed line corresponds to the threshold used to determine if open or closed. The circle radius is proportional to the population transferred during the spawning event and separated into eq-G- (teal), eq-T (purple), and eq-G+ (olive). (bottom) The $S_1$ population decay for eq-G-, eq-T, and eq-G+ for the first ps after photoexcitation.

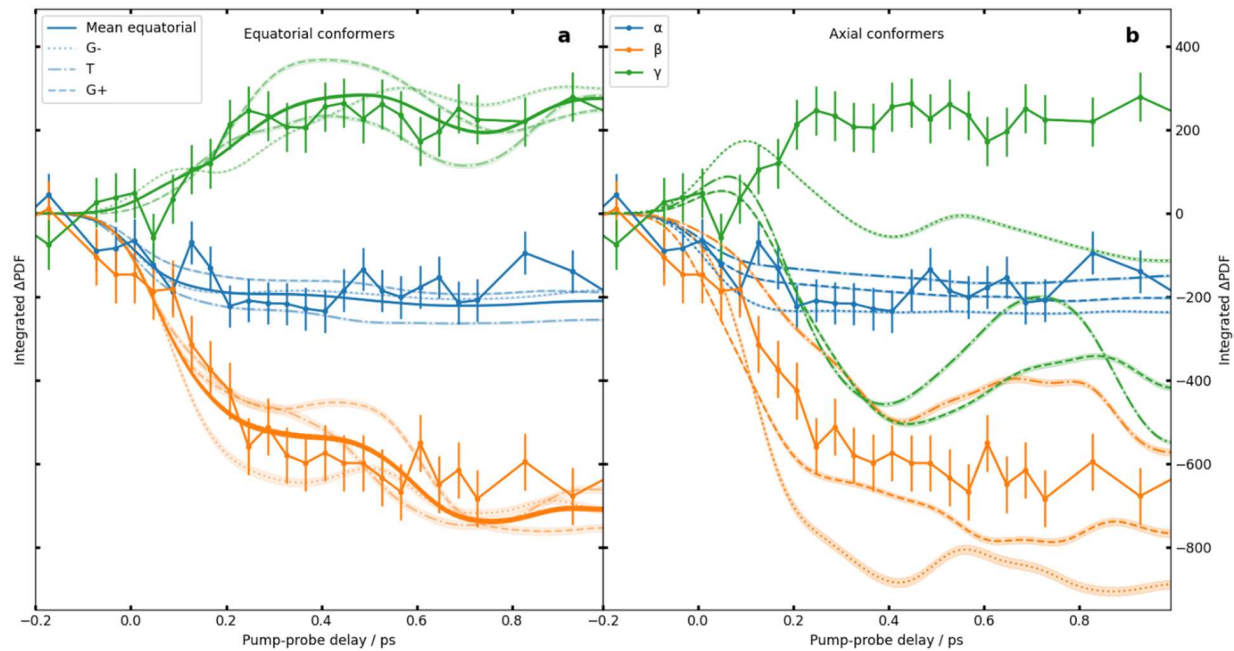

**Fig. S3.** Experimental and Simulated Signal Rise-Times Along Coordination Shells. (a) Temporal evolution of the integrated regions of the first (α, blue), second (β, orange) and third (γ, green) coordination shells of the experimental (connected points with error bars) and simulated ΔPDF averaged over the three equatorial rotamers (continuous lines). Simulated ΔPDF signals of the individual equatorial rotamers are shown as dotted (G-), dash-dotted (T), and dashed (G+) lines. The experimental data show quantitative agreement with the average of the simulated equatorial rotamer signals. (b) Analogous comparison between the experimental and simulated ΔPDF signals of the individual axial rotamers. Error bars represent a 68% confidence interval obtained from bootstrap analysis. The error bars of the simulations are visualized by the width of the lines and reflect convergence with respect to initial condition sampling.

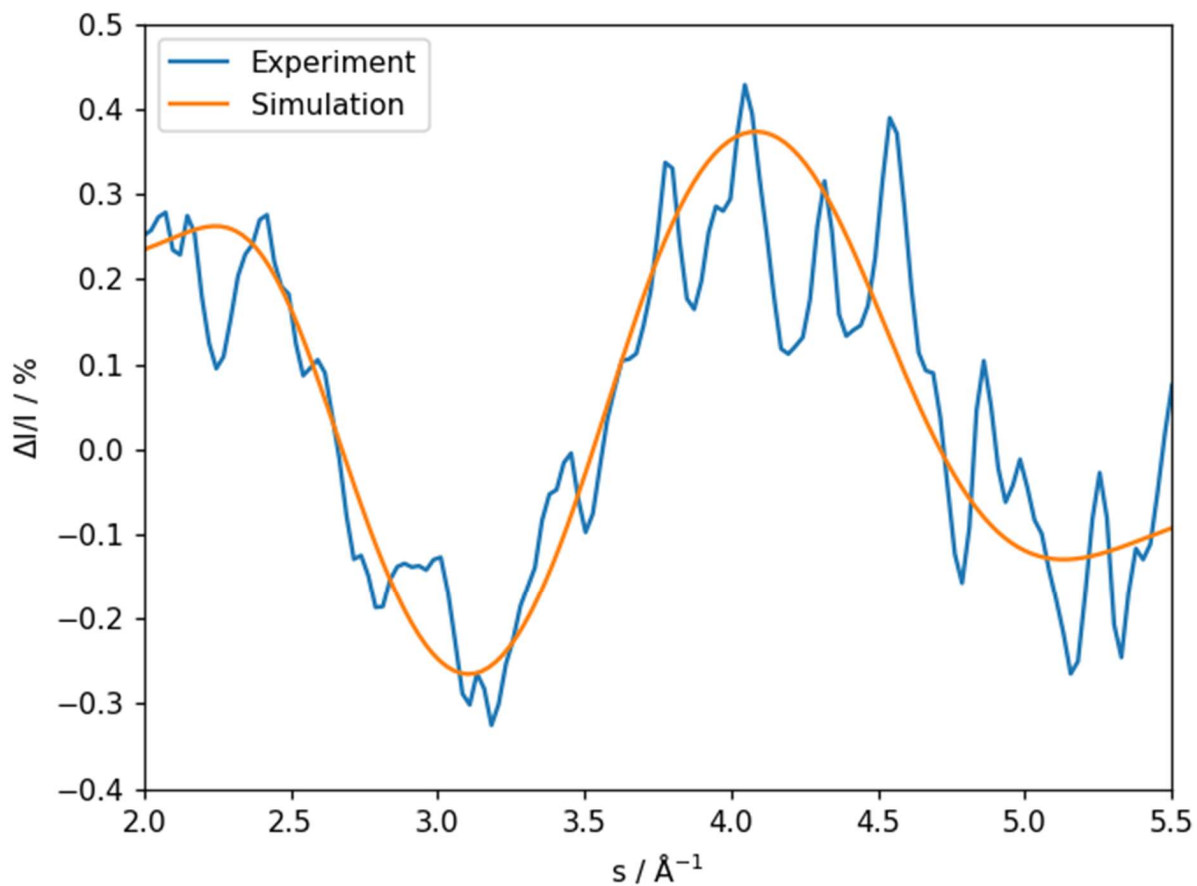

**Fig. S4.** Experimental and Simulated Relative Difference Diffraction Signals. Comparison between experimental and simulated relative difference diffraction signals (I(1 ps)-I(steady-state))/I(steady-state). The simulation is scaled by a factor of 0.025 to match the amplitude of the experiment. This suggests an excitation ratio of 2.5 %.

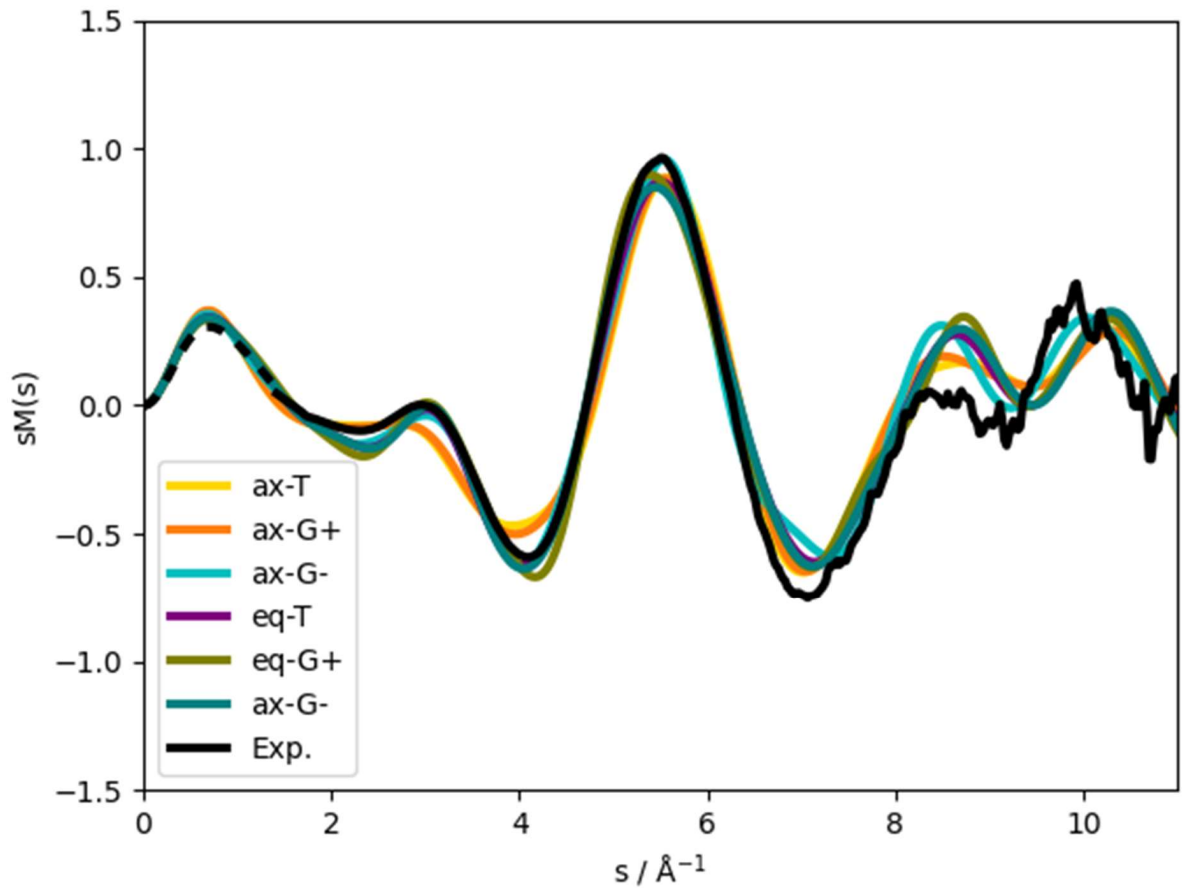

**Fig. S5.** Comparison of the experimental modified molecular diffraction sM(s) with simulations of all six phellandrene conformers. The experimental sM(s) is replaced by the average of the simulated sM(s) below 1.5 Å$^{-1}$ to avoid artifacts from the hole in the detector in the experimental PDF (see Fig. 2a). As in Fig. 2a, we find good agreement between the experimental sM(s) and four of the six possible conformers, all three equatorial and the ax-G- conformer, in a wide range from 1.5 Å$^{-1}$ to 8 Å$^{-1}$. The agreement in Fig. 2a is worse in the range >8 Å$^{-1}$, which might be due to shortcomings in the subtraction of the atomic scattering background and limited signal-to-noise levels. The ax-T and ax-G+ conformers show significantly worse agreement with the experimental data in the region between 2 Å$^{-1}$ and 4.5 Å$^{-1}$.

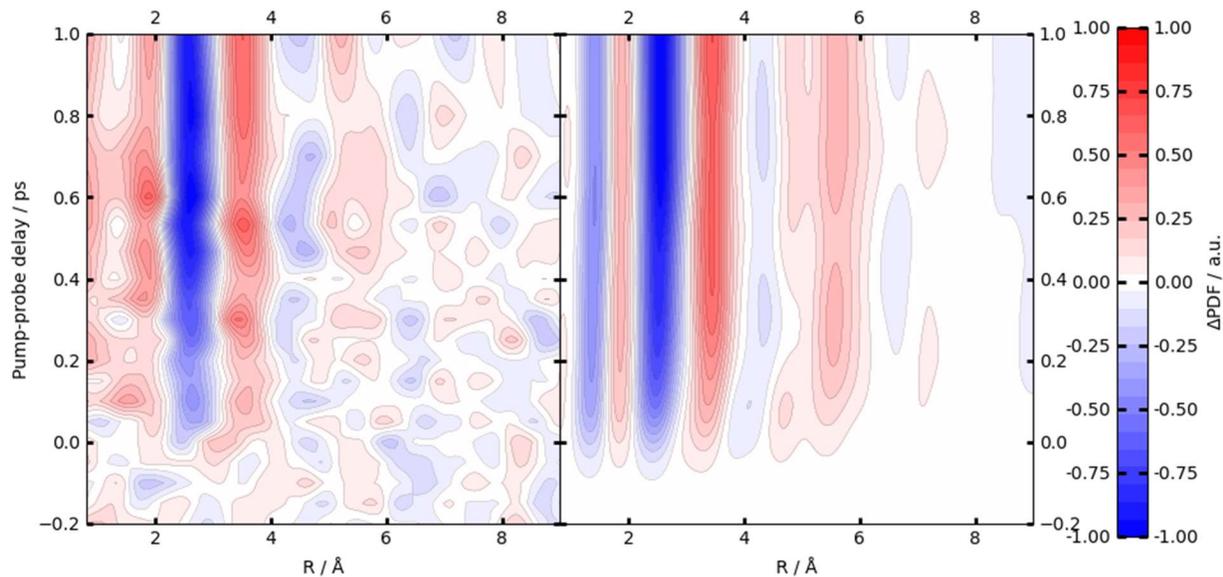

**Fig. S6.** Experimental ΔPDF(r,t) (left) and the average of the simulated ΔPDF(r,t) of equatorial rotamers (right) . Both ΔPDF(r,t) are generated while setting the s range <0.6 Å$^{-1}$ to zero to assess the bias to the experimental ΔPDF(r,t) by treatment of the s<0.6 Å$^{-1}$ for Fig. 3 (see supplementary note 4). The main artifacts introduced by setting s<0.6 Å$^{-1}$ to zero are a smooth positive contribution to distances <2 Å and a negative contribution to distances > 4 Å.

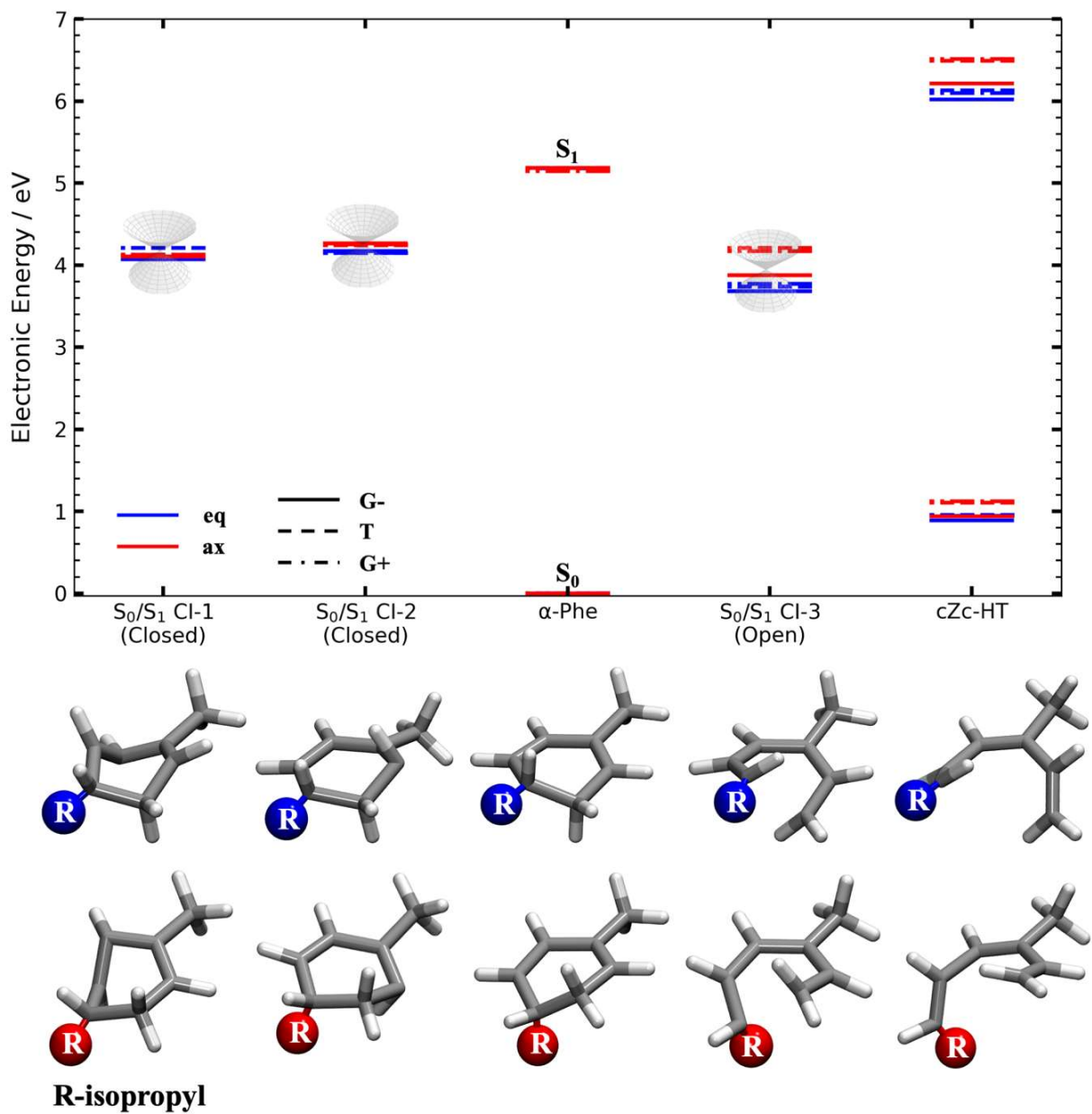

**Fig. S7.** Critical Points Along Nonradiative Relaxation Pathways in Ax and Eq αPH. The computed ax (red) and eq(blue) PESs at critical points along the OOP and ring-opening reaction coordinates in αPH. The energies are relative to each isomer's respective $S_0$ minimum ground-state energy. Geometries are shown in the bottom of the figure with the red (ax) and blue (eq) spheres representing the location of the ISO. Computed at the α(0.82)-SA2-CAS(6,4)-SCF/6-31G* level of theory. See **Table S2** for coordinates, energies, and CI vectors.

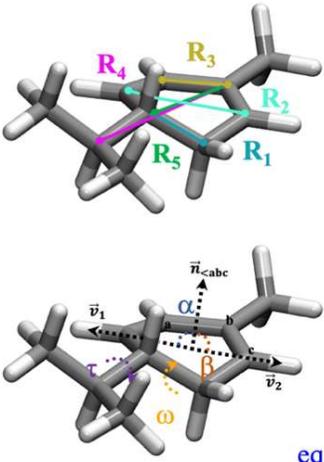

**Fig. S8.** The Structural Properties of the Critical Points. Computed structural properties of the critical points for all eq and ax isomers at the α(0.82)-SA2-CAS(6,4)-SCF/6-31G* level of theory. See Table S2 for coordinates, energies, and CI vectors.

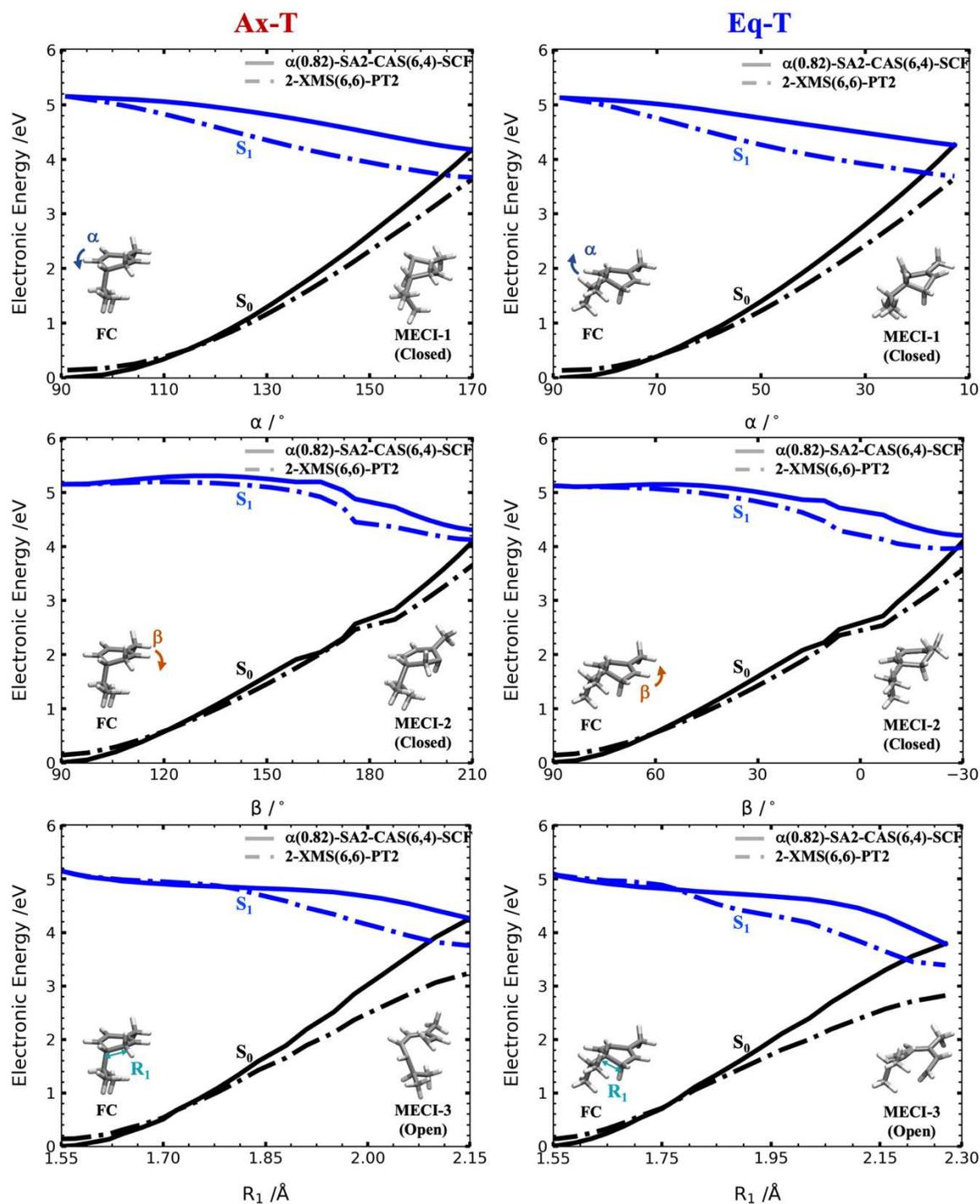

**Fig. S9.** Benchmarking α-SA-CASSCF Against XMSPT2 on $S_1$ with the G- Ax and Eq Conformers. Potential energy surface scans along the OOP (CI-1 and CI-2) and ring-opening (CI-3) coordinates on the $S_1$ electronic state. The pathways were generated from geodesic interpolation between the FC point and CI-1, CI-2, and CI-3 MECI structures, respectively, optimized at the α(0.82)-SA2-CAS(6,4)SCF/6-31G* level of theory and compared against single-point energy calculations with SA2-XMS-CAS(6,6)-PT2/6-31G*.

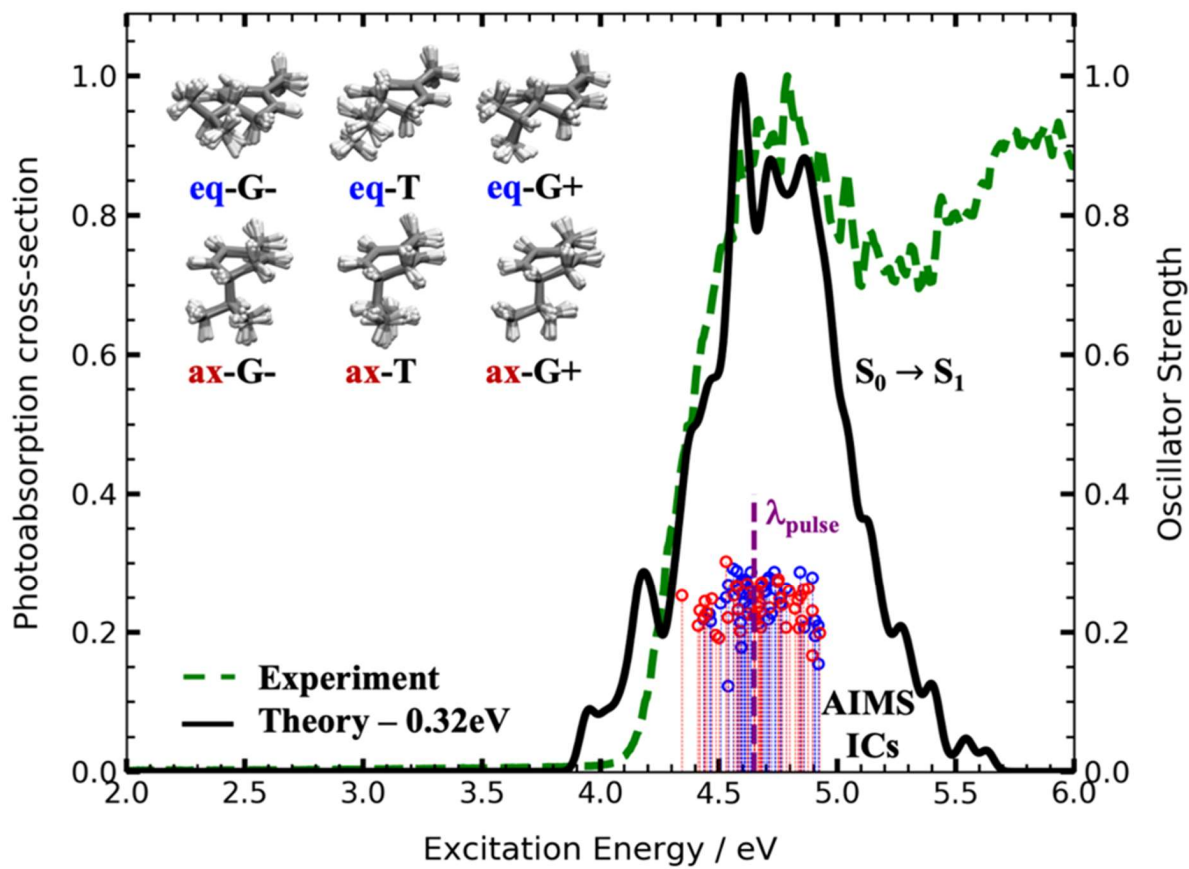

**Fig. S10.** UV Electronic Absorption Spectra of αPH: The UV electronic absorption spectrum was generated from 600 initial conditions sampled from a ground-state harmonic Wigner distribution. The AIMS dynamics simulations used 15 initial conditions for each conformer. The energy and oscillator strength for each of the initial conditions (randomly sampled with the restriction that they were within 0.3eV of the pump pulse energy used in the UED experiment) are shown with red/blue vertical lines for the axial/equatorial initial conditions respectively. The inset shows the starting geometries for each conformer.

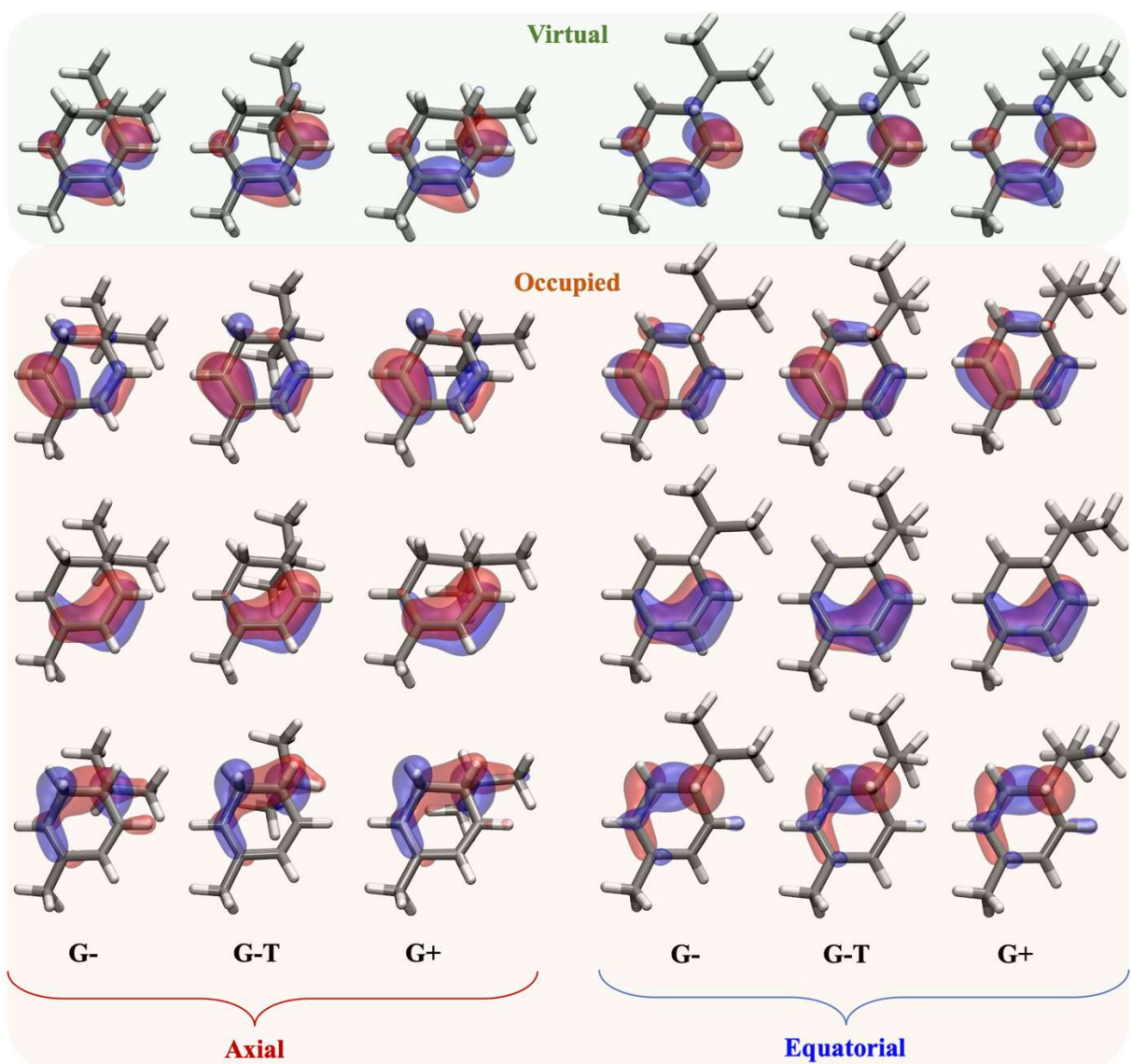

**Fig. S11.** SA-CASSCF Natural MOs at the $S_0$ Minima for all six aPH Isomers. The α(0.82)-SA2-CAS(6,4)SCF natural orbitals for the critical points along the α, β, and γ Z/E photoisomerization pathways. Blue and red correspond to 0.05 and -0.05 e-/Å$^3$ isovalues, respectively.

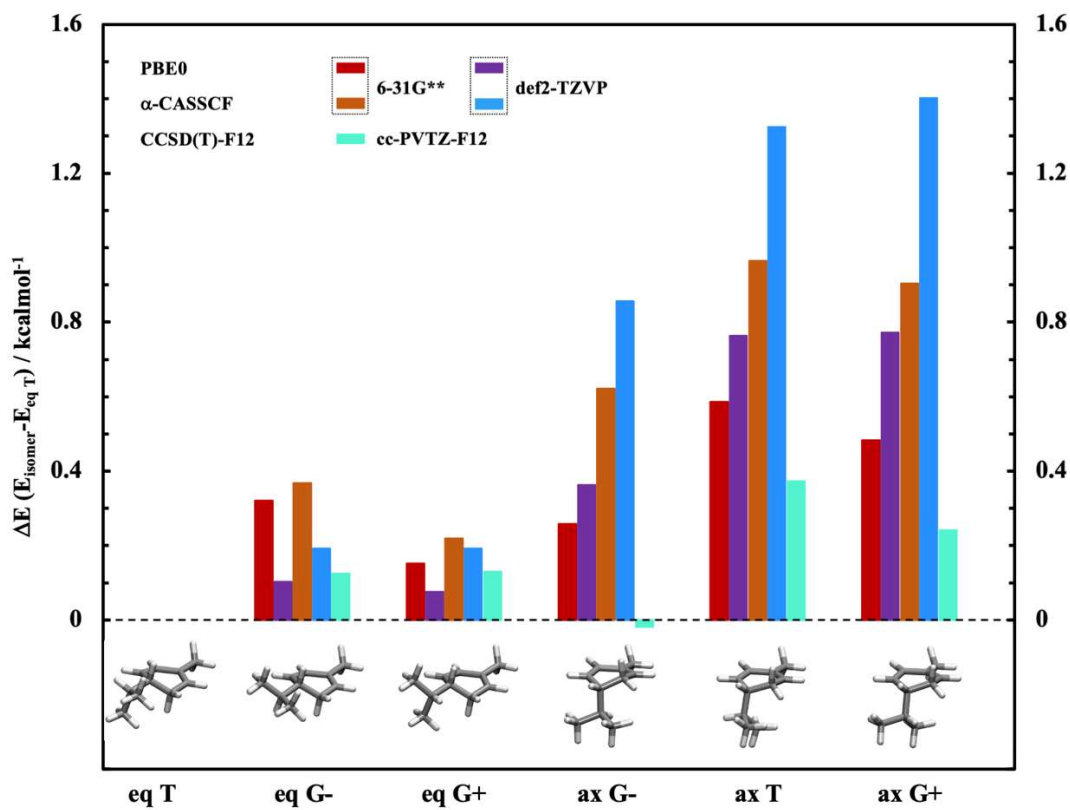

| Method | Basis Set | ΔE ($E_{isomer} - E_{eq\,T}$) / kcalmol$^{-1}$ | | | | | |
|---|---|---|---|---|---|---|---|
| | | eq T | eq G- | eq G+ | ax G- | ax T | ax G+ |
| PBE0 | 6-31G** | 0.00 | 0.32 | 0.15 | 0.26 | 0.59 | 0.48 |
| | def2-TZVP | 0.00 | 0.10 | 0.08 | 0.36 | 0.76 | 0.77 |
| a-CASSCF | 6-31G** | 0.00 | 0.37 | 0.22 | 0.62 | 0.96 | 0.90 |
| | def2-TZVP | 0.00 | 0.19 | 0.19 | 0.86 | 1.32 | 1.40 |
| CCSD(T)-F12 | cc-pVDZ-F12 | 0.00 | 0.12 | 0.13 | -0.02 | 0.37 | 0.24 |

**Fig. S12.** The relative energies of the S$_0$ minima of the six most stable aPH isomers optimized at the PBE0/6-31G** level of theory and single-point energies computed with their labeled methods. a-CASSCF corresponds to a(0.82)-SA2-CAS(6,4)SCF. The aPH conformers are shown below their computed relative energies.

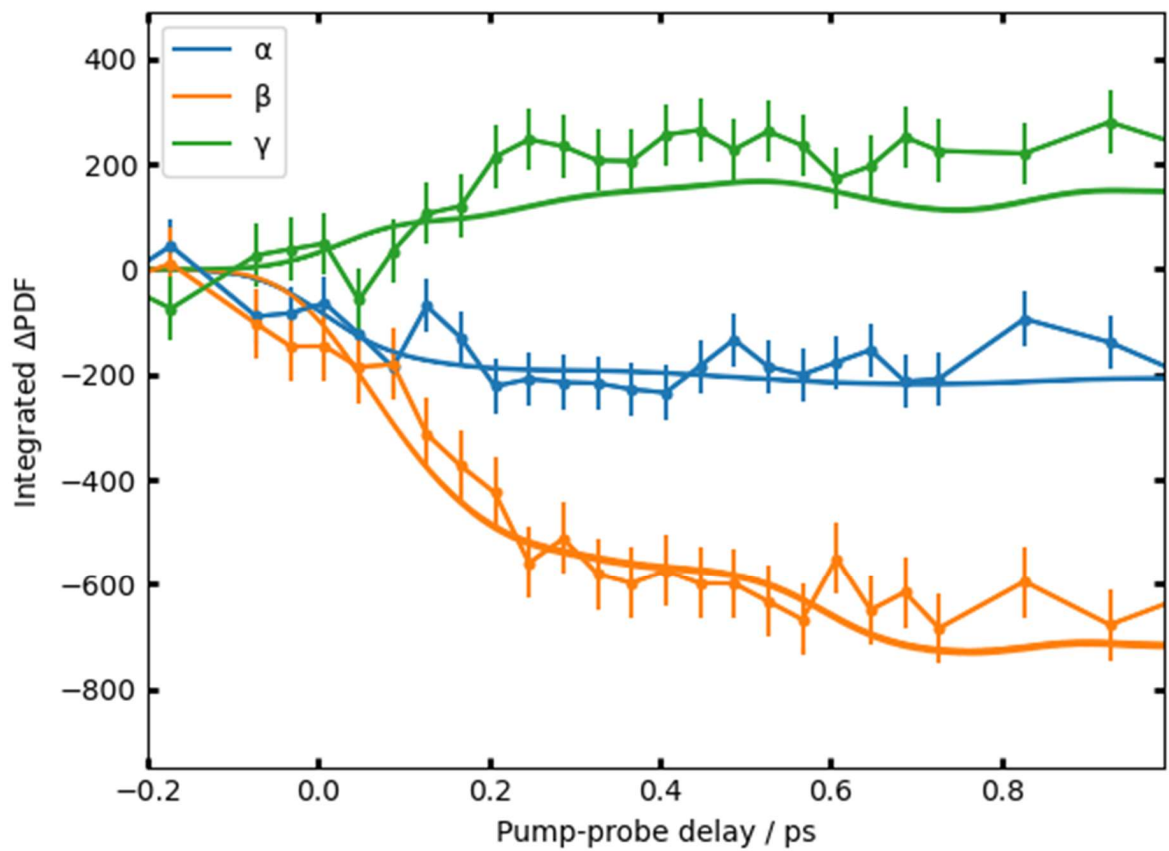

**Fig. S13.** Comparison of the experimental data to a linear combination of 80% eq-αPH and 20% ax-αPH, analogous to Fig. 3d. The intensity of the α and β features is not substantially changed by the linear combination, since both the eq-αPH and ax-αPH conformers open the ring. With a 20 % contribution from ax-αPH, the intensity of the γ-signature reduced far enough (see Fig. S3) that the simulation is outside the error bars (68% confidence interval) of the experiment.

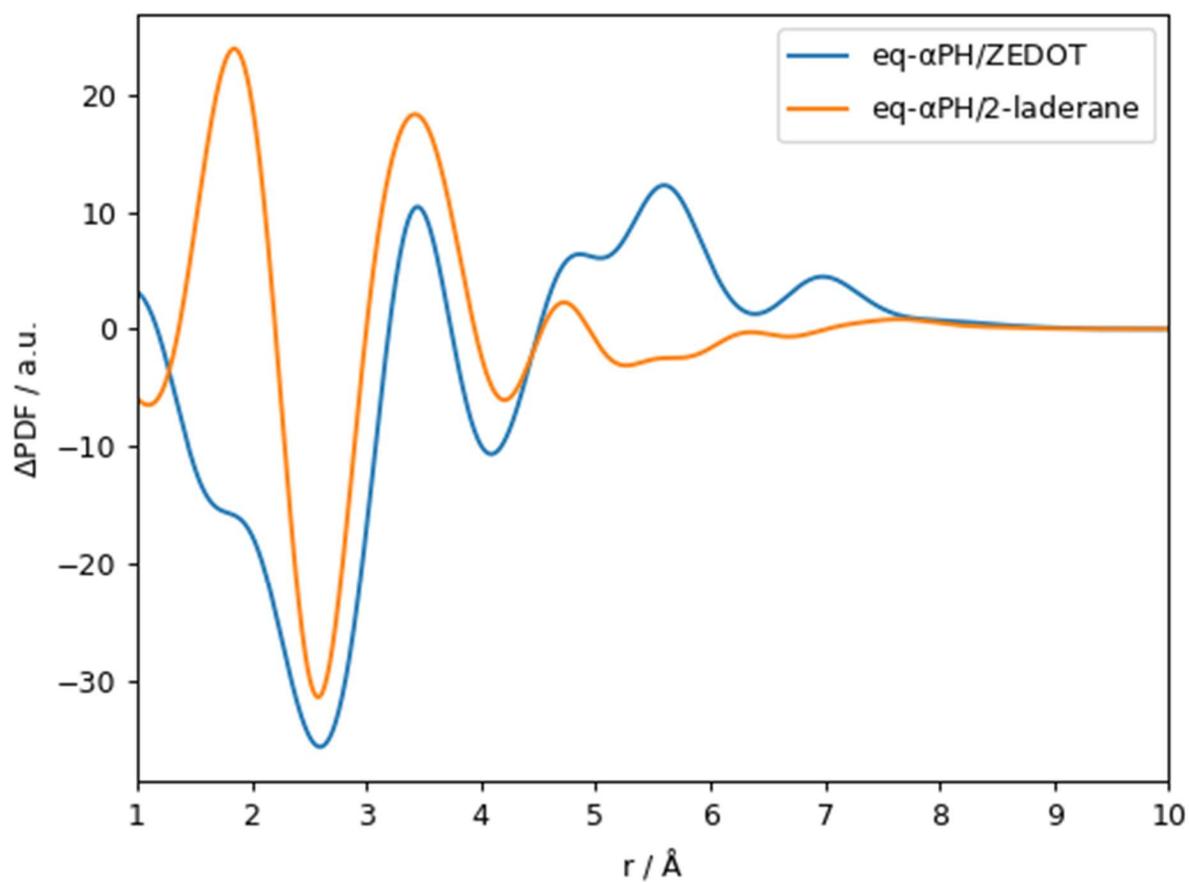

**Fig. S14.** Comparison of simulated ΔPDF signatures of the ZEDOT and 2-ladderane photoproducts. The ΔPDF signatures are generated analogous to **Fig. 2**.

## Supplementary Tables

| Isomer | CI-1 (%) | | CI-2 (%) | | CI-3 (%) | |
|---|---|---|---|---|---|---|
| | αPH | ZZ/ZE-DOT | αPH | ZZ/ZEDOT | αPH | ZZ/ZE-DOT |
| **Axial** | 6 +/- 2 | 0 | 46 +/- 5 | 0 | 24 +/- 4 | 24 +/- 4 |
| G- | 0 | 0 | 13 +/- 6 | 0 | 50 +/- 7 | 37 +/- 6 |
| T | 12 +/- 6 | 0 | 71 +/- 8 | 0 | 3 +/- 1 | 14 +/- 6 |
| G+ | 7 +/- 3 | 0 | 55 +/- 9 | 0 | 17 +/- 5 | 21 +/- 7 |
| **Equatorial** | 5 +/- 2 | 0 | 35 +/- 4 | 0 | 30 +/- 3 | 30 +/- 3 |
| G- | 9 +/- 4 | < 1 | 22 +/- 6 | 0 | 38 +/- 7 | 31 +/- 6 |
| T | 6 +/- 2 | 0 | 34 +/- 8 | 0 | 30 +/- 6 | 30 +/- 6 |
| G+ | 0 | 0 | 50 +/- 8 | 0 | 21 +/- 5 | 29 +/- 6 |

**Table S1.** Computed Quantum Yield for Eq and Ax aPH. The computed quantum yields for ax (red) and eq (blue) aPH from the AIMS simulation. Errors represent a 68% confidence interval and obtained from bootstrap analysis. The geometries associated with each CI are included in the supplementary structures.

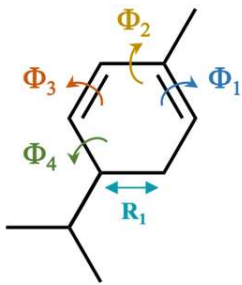

| Conformer | $R_1$ (Å) | $|\Phi_1|$ (°) | $|\Phi_2|$ (°) | $|\Phi_3|$ (°) | $|\Phi_4|$ (°) |
|---|---|---|---|---|---|
| αPH | ≤ 1.8 | ≤ 80 | ≤ 80 | ≤ 80 | - |
| cZc DOT | > 1.8 | ≤ 80 | ≤ 80 | ≤ 80 | - |
| cZt/tZc DOT | > 1.8 | ≤ 80 or ≥ 100 | ≤ 80 | ≥ 100 or ≤ 80 | - |
| tZt DOT | > 1.8 | ≥ 100 | ≤ 80 | ≥ 100 | - |
| ZZDOT | > 1.8 | ≥ 100 | ≤ 80 | ≥ 100 | ≤ 80 |
| ZEDOT | > 1.8 | ≥ 100 | ≤ 80 | ≥ 100 | ≥ 100 |

**Table S2.** Binning criteria for aPH conformers upon relaxation to the ground-electronic state. Each row corresponds to a specific conformer along with the structural parameters used to classify a specific geometry R(t). The carbon-carbon distance ($R_1$), and the absolute values of the four dihedral angles ($F_1$, $F_2$, $F_3$, and $F_4$) are shown above on the aPH structure.

| | **ax-G- S₀ Minimum (FC)** | | | |
|---|---|---|---|---|
| Cartesian coordinates / Å | 26 | | | |
| | C | 0.5603236327 | 0.4023961588 | -0.7534127770 |
| | C | 0.7083580082 | 0.2280638902 | -2.2917607550 |
| | C | -0.4806859329 | 1.4601042097 | -0.3427323697 |
| | C | 0.1905270201 | -0.9037782124 | -0.0835836769 |
| | C | 1.1561432757 | 1.5217719416 | -2.9836108245 |
| | C | 1.6972509320 | -0.8934076400 | -2.6346771681 |
| | C | -1.8966775403 | 0.9530193094 | -0.4431649948 |
| | C | -2.1892619383 | -0.3262247477 | -0.1896019727 |
| | C | -1.1016287345 | -1.2366900564 | 0.1308329124 |
| | C | -3.5971279122 | -0.8637417517 | -0.1983626931 |
| | H | 1.5295358649 | 0.7243728535 | -0.3774203022 |
| | H | -0.2652108994 | -0.0513786141 | -2.6868248926 |
| | H | -0.3612853868 | 2.3725422211 | -0.9128206563 |
| | H | -0.2910190030 | 1.7300974998 | 0.6974986094 |
| | H | 0.9717223103 | -1.6003674665 | 0.1631218703 |
| | H | 2.0942398065 | 1.8821564697 | -2.5675740747 |
| | H | 1.3142229773 | 1.3480004559 | -4.0437769645 |
| | H | 0.4259905321 | 2.3174470212 | -2.8965062503 |
| | H | 2.6834867910 | -0.6825871333 | -2.2264033648 |
| | H | 1.8034611151 | -0.9919160193 | -3.7107171621 |
| | H | 1.3735024045 | -1.8532756844 | -2.2505699153 |
| | H | -2.6788309627 | 1.6577899201 | -0.6692289336 |
| | H | -1.3468031635 | -2.2011433469 | 0.5430824664 |
| | H | -4.3167380410 | -0.0869263586 | -0.4286740953 |
| | H | -3.8554407042 | -1.2860978305 | 0.7695949806 |
| | H | -3.7080644517 | -1.6572470888 | -0.9328169955 |

| | |
|---|---|
| $S_0$ energy / H | -387.96054165320811 |
| $S_0$ CI eigenvector | -0.97504331747285  X36 X37 X38 |
| | 0.11393573558125  X36 X37 X39 |
| | 0.09694743023016  X36 X37 A38 B39 |
| | 0.09694743023016  X36 X37 B38 A39 |
| | 0.08914326689277  X36 X38 X39 |
| | -0.05944949541773  X36 A37 B38 X39 |
| | -0.05944949541773  X36 B37 A38 X39 |
| | -0.02940805078553  A36 X37 B38 X39 |
| | -0.02940805078553  B36 X37 A38 X39 |
| $S_1$ energy / H | -387.77196192146300 |
| $S_1$ CI eigenvector | 0.66569875346325  X36 X37 A38 B39 |
| | 0.66569875346325  X36 X37 B38 A39 |
| | 0.16659600268016  X36 A37 X38 B39 |
| | 0.16659600268016  X36 B37 X38 A39 |
| | 0.14470360526893  X36 X37 X39 |
| | 0.13321696755136  X36 X37 X38 |
| | -0.10551863283796  X36 X38 X39 |
| | 0.04302315766171  X36 A37 B38 X39 |
| | 0.04302315766171  X36 B37 A38 X39 |
| | -0.03902941742766  A36 X37 B38 X39 |
| | -0.03902941742766  B36 X37 A38 X39 |
| | 0.02509392450520  A36 X37 X38 B39 |
| | 0.02509392450520  B36 X37 X38 A39 |

**ax-G- $S_1$/$S_0$ MECI-1 (Closed)**

| | | | | |
|---|---|---|---|---|
| Cartesian coordinates / Å | 26 | | | |
| | C | 0.1621130953 | 0.3955961655 | -1.1141988442 |
| | C | 0.7554693358 | 0.4585408063 | -2.5301436349 |

|   |   |   |   |
|---|---|---|---|
| C | -0.9982808394 | 1.3903029991 | -0.8564786760 |
| C | -0.5200554079 | -0.8934853963 | -0.7580407137 |
| C | 1.4207136966 | 1.8157466977 | -2.7849453037 |
| C | 1.7662869611 | -0.6692895249 | -2.7708141217 |
| C | -2.2081768506 | 0.6026907689 | -0.4024733111 |
| C | -2.1335035690 | -0.4945820310 | 0.4651025289 |
| C | -0.9020356518 | -1.1433888871 | 0.6157552912 |
| C | -3.4055396165 | -1.0750878568 | 1.0361741662 |
| H | 0.9494501058 | 0.5417641763 | -0.3790014621 |
| H | -0.0620339365 | 0.3377675863 | -3.2405277112 |
| H | -1.2455509581 | 1.8841126506 | -1.7897108748 |
| H | -0.7295990292 | 2.1744307534 | -0.1492908892 |
| H | -0.7957838139 | -1.5654818326 | -1.5618515905 |
| H | 2.2390688093 | 1.9789179397 | -2.0885236535 |
| H | 1.8288636695 | 1.8626898754 | -3.7897525351 |
| H | 0.7259965935 | 2.6411199744 | -2.6750474642 |
| H | 2.6222121244 | -0.5717508947 | -2.1085328887 |
| H | 2.1334163496 | -0.6449568693 | -3.7922160811 |
| H | 1.3351328597 | -1.6522678531 | -2.6068862224 |
| H | -3.1783057381 | 1.0133707489 | -0.6255953245 |
| H | -0.8978478290 | -2.1402797112 | 1.0182670790 |
| H | -3.8378633989 | -1.7899590253 | 0.3425576005 |
| H | -4.1407189015 | -0.2999167599 | 1.2253504958 |
| H | -3.1938291008 | -1.5950621513 | 1.9626938446 |

| | |
|---|---|
| S$_0$ energy / H | -387.80784501018326 |
| S$_0$ CI eigenvector | 0.64709438890734  X36 X37 A38 B39 |
| | 0.64709438890734  X36 X37 B38 A39 |
| | 0.34695217065350  X36 X37 X38 |
| | -0.12074460502525  X36 A37 B38 X39 |
|  | -0.12074460502525  X36 B37 A38 X39 |
|  | 0.07475597349097  X36 A37 X38 B39 |
|  | 0.07475597349097  X36 B37 X38 A39 |
|  | -0.03931498098687  X36 X38 X39 |
| S$_1$ energy / H | -387.80784494827003 |
| S$_1$ CI eigenvector | 0.91822500256939  X36 X37 X38 |
|  | -0.24880197931689  X36 X37 A38 B39 |
|  | -0.24880197931689  X36 X37 B38 A39 |
|  | -0.09515242037606  X36 X38 X39 |
|  | 0.08644154315953  X36 A37 X38 B39 |
|  | 0.08644154315953  X36 B37 X38 A39 |
|  | 0.05692936396372  X36 A37 B38 X39 |
|  | 0.05692936396372  X36 B37 A38 X39 |
|  | -0.03993705200290  X37 X38 X39 |
|  | -0.02522895954515  X36 X37 X39 |

## ax-G- S$_1$/S$_0$ MECI-2 (Closed)

| Cartesian coordinates / Å | 26 |  |  |  |
|---|---|---|---|---|
|  | C | 0.2778394844 | 0.7402758905 | -0.9436995820 |
|  | C | 0.6503719408 | 0.2715291507 | -2.3871127407 |
|  | C | -1.0396062394 | 1.5584442046 | -0.8519567648 |
|  | C | 0.1842609901 | -0.4772280282 | -0.0541691708 |
|  | C | 0.6567092825 | 1.4346682168 | -3.3863680272 |
|  | C | 2.0183001147 | -0.4274011723 | -2.4178296436 |
|  | C | -2.3414841531 | 0.7087306843 | -0.8612023319 |
|  | C | -2.2141956516 | -0.3553061164 | 0.0641837175 |

|   |                |                |                |
|---|---------------:|---------------:|---------------:|
| C | -0.9600545966  | -1.0705232758  |  0.3100330113  |
| C | -3.4061451847  | -0.7973599899  |  0.8573050918  |
| H |  1.1179291181  |  1.3429610916  | -0.5972064999  |
| H | -0.0958012795  | -0.4474428770  | -2.7161738790  |
| H | -1.0688533166  |  2.3007857938  | -1.6353233566  |
| H | -1.0285218047  |  2.1194928466  |  0.0803919283  |
| H |  1.1078654787  | -0.9194881186  |  0.2793939825  |
| H |  1.2757531546  |  2.2543892213  | -3.0301579862  |
| H |  1.0621302520  |  1.1067821862  | -4.3384028664  |
| H | -0.3369049217  |  1.8191988087  | -3.5744596838  |
| H |  2.8050280313  |  0.2412133230  | -2.0767374427  |
| H |  2.2643399127  | -0.7320552651  | -3.4299009112  |
| H |  2.0493782122  | -1.3194550302  | -1.8017253245  |
| H | -1.9867231265  | -0.0617963457  | -1.6329539545  |
| H | -0.9436437217  | -1.9469927486  |  0.9350040700  |
| H | -3.9033571669  |  0.0926770406  |  1.2255139528  |
| H | -3.1809887850  | -1.4927685603  |  1.6583762018  |
| H | -4.0991488352  | -1.2694627991  |  0.1673533469  |

| | |
|---|---|
| $S_0$ energy / H | -387.80197661779243 |
| $S_0$ CI eigenvector |  0.61288824429786  X36 X37 A38 B39 |
| |  0.61288824429786  X36 X37 B38 A39 |
| | -0.47210234524164  X36 X37 X38 |
| | -0.07823966371266  X36 A37 B38 X39 |
| | -0.07823966371266  X36 B37 A38 X39 |
| |  0.05200492398262  A36 X37 B38 X39 |
| |  0.05200492398262  B36 X37 A38 X39 |
| | -0.05053978461282  X36 A37 X38 B39 |
| | -0.05053978461282  X36 B37 X38 A39 |
| |  0.03900987873768  X36 X38 X39 |

|   |   |   |
|---|---|---|
| | 0.03798693530241 | X37 X38 X39 |
| S$_1$ energy / H | -387.80197656569987 | |
| S$_1$ CI eigenvector | -0.86047987680726 | X36 X37 X38 |
| | -0.33611040007945 | X36 X37 A38 B39 |
| | -0.33611040007945 | X36 X37 B38 A39 |
| | -0.09283438295822 | X36 A37 X38 B39 |
| | -0.09283438295822 | X36 B37 X38 A39 |
| | 0.07457338036087 | X36 X38 X39 |
| | 0.05910489653625 | X37 X38 X39 |
| | 0.03898805618125 | X36 A37 B38 X39 |
| | 0.03898805618125 | X36 B37 A38 X39 |
| | -0.03054638312189 | A36 X37 B38 X39 |
| | -0.03054638312189 | B36 X37 A38 X39 |
| | 0.03000672656738 | A36 X37 X38 B |

**ax-G- S$_1$/S$_0$ MECI-3 (Open)**

Cartesian coordinates / Å

| | | | |
|---|---|---|---|
| C | 0.8417220268 | 0.3094150283 | -0.7527754442 |
| C | 0.6684700878 | 0.2214409531 | -2.2643079427 |
| C | -0.6958979840 | 1.6461909802 | -0.1426474445 |
| C | 0.2844862496 | -0.6297399299 | 0.1341828001 |
| C | 0.9992342452 | 1.5374316838 | -2.9762029416 |
| C | 1.5616871132 | -0.9016635253 | -2.8201445510 |
| C | -1.8763693304 | 1.0561312980 | -0.6336239893 |
| C | -2.1094764742 | -0.3083304716 | -0.3426723474 |
| C | -1.0710316514 | -1.1015591199 | 0.1157619641 |
| C | -3.5023331242 | -0.8672701793 | -0.4822571987 |
| H | 1.8136285935 | 0.6807553281 | -0.4577909945 |
| H | -0.3608848334 | -0.0334349404 | -2.4824666315 |

|   |   |   |   |   |
|---|---|---|---|---|
| H | -0.3559649374 | 2.5761288705 | -0.5647945372 |
| H | -0.4631322682 | 1.5528539700 | 0.9031165436 |
| H | 0.8851773650 | -0.9178316287 | 0.9866735061 |
| H | 2.0128804599 | 1.8602113593 | -2.7513681460 |
| H | 0.9263962167 | 1.4154061612 | -4.0529847435 |
| H | 0.3226519642 | 2.3354558845 | -2.6941835457 |
| H | 2.6122420179 | -0.6942201329 | -2.6317535898 |
| H | 1.4349543141 | -1.0013416331 | -3.8953038697 |
| H | 1.3229835357 | -1.8563926264 | -2.3635138377 |
| H | -2.4347179111 | 1.5415274257 | -1.4141879774 |
| H | -1.2872180516 | -2.0506155676 | 0.5759124282 |
| H | -4.0695445146 | -0.6663727822 | 0.4229638103 |
| H | -3.4986832287 | -1.9374193803 | -0.6527548730 |
| H | -4.0278662559 | -0.3845268228 | -1.2992176941 |

| | |
|---|---|
| $S_0$ energy / H | -387.81621596389630 |
| $S_0$ CI eigenvector | -0.68422866365199  X36 X37 A38 B39 |
| | -0.68422866365199  X36 X37 B38 A39 |
| | -0.20802236504258  X36 X37 X38 |
| | 0.06916271094306  X36 A37 B38 X39 |
| | 0.06916271094306  X36 B37 A38 X39 |
| | 0.06569307897616  A36 X37 B38 X39 |
| | 0.06569307897616  B36 X37 A38 X39 |
| | 0.03658466309883  X36 X38 X39 |
| $S_1$ energy / H | -387.81621591282703 |
| $S_1$ CI eigenvector | 0.95912177525701  X36 X37 X38 |
| | -0.14830333425409  X36 X37 A38 B39 |
| | -0.14830333425409  X36 X37 B38 A39 |
| | -0.12097924030911  X37 X38 X39 |

-0.10079638786795  X36 X38 X39

0.06150863043967  X36 X37 X39

0.04052344337378  X36 A37 X38 B39

0.04052344337378  X36 B37 X38 A39

0.03600610810151  A36 X37 X38 B39

0.03600610810151  B36 X37 X38 A39

-0.02612209225221  A36 B37 X38 X39

-0.02612209225221  B36 A37 X38 X39

## **ax-G- cZc-ZDOT Minimum**

Cartesian coordinates / Å     26

| | | | |
|---|---|---|---|
| C | 1.3090863484 | -0.5878350482 | -0.7434429896 |
| C | 1.0465692236 | -0.7793383837 | -2.2162131453 |
| C | -1.1432254930 | 2.0674572329 | -0.3657589622 |
| C | 0.4441091090 | -0.6690636364 | 0.2553290459 |
| C | 1.4842781081 | 0.4594064191 | -3.0084413775 |
| C | 1.7748265009 | -2.0329293034 | -2.7220502343 |
| C | -1.9719833495 | 1.0816773488 | -0.7649518194 |
| C | -2.0283354018 | -0.2931430453 | -0.2757166191 |
| C | -0.9959479244 | -1.0165419993 | 0.1787065913 |
| C | -3.4109996727 | -0.9102520116 | -0.3126350699 |
| H | 2.3385346354 | -0.3788405082 | -0.4902012134 |
| H | -0.0175593621 | -0.9206312376 | -2.3640545539 |
| H | -1.2235225903 | 3.0521754435 | -0.7893540989 |
| H | -0.3877864618 | 1.9206509692 | 0.3811859555 |
| H | 0.8284274383 | -0.5527355098 | 1.2586502807 |
| H | 2.5488757139 | 0.6441614973 | -2.8867486787 |
| H | 1.2924245572 | 0.3254287432 | -4.0691429647 |
| H | 0.9497915720 | 1.3424698104 | -2.6778513910 |
| H | 2.8491079144 | -1.9447799775 | -2.5808029759 |
| H | 1.5920944455 | -2.1841464761 | -3.7820969853 |
| H | 1.4415587258 | -2.9196534359 | -2.1927677671 |
| H | -2.7421950526 | 1.3400666511 | -1.4754598184 |
| H | -1.2265331534 | -1.9934948160 | 0.5762032124 |
| H | -4.1086180137 | -0.3348258987 | 0.2898151122 |
| H | -3.4070753781 | -1.9303711735 | 0.0509286506 |
| H | -3.7990724391 | -0.9180916543 | -1.3282781840 |

$S_0$ energy / H    -387.92413487437199

$S_0$ CI eigenvector    
  0.97532175301649 X36 X37 X38  
-0.09516964682509 X36 X37 X39  
-0.09126436593515 X36 X37 A38 B39  
-0.09126436593515 X36 X37 B38 A39  
-0.08061804676734 X36 X38 X39  
-0.06963203413384 X36 A37 B38 X39  
-0.06963203413384 X36 B37 A38 X39  
-0.03808325004110 A36 B37 X38 X39  
-0.03808325004110 B36 A37 X38 X39  
-0.03659396191285 X37 X38 X39  
-0.03100542062448 X36 A37 X38 B39  
-0.03100542062448 X36 B37 X38 A39  

$S_1$ energy / H    -387.73029552375090

$S_1$ CI eigenvector    
-0.65810289146586 X36 X37 A38 B39  
-0.65810289146586 X36 X37 B38 A39  
  0.16560945436037 X36 A37 X38 B39  
  0.16560945436037 X36 B37 X38 A39  
-0.15734002442813 X36 X37 X39

-0.11706601371077  X36 X37 X38

0.09710454236041  A36 X37 X38 B39

0.09710454236041  B36 X37 X38 A39

0.08284563231944  X36 X38 X39

0.06809940224884  A36 B37 X38 X39

0.06809940224884  B36 A37 X38 X39

0.04489481348085  A36 X37 B38 X39

0.04489481348085  B36 X37 A38 X39

0.03636524577451  X37 X38 X39

---

## ax-T S$_0$ Minimum (FC)

| Cartesian coordinates / Å | 26 | | | |
|---|---|---|---|---|
| | C | 0.4502477944 | 0.3858873308 | -0.8255864458 |
| | C | 0.7687565070 | 0.2731381052 | -2.3441796670 |
| | C | -0.6076555934 | 1.4488225252 | -0.4521150388 |
| | C | 0.0835415899 | -0.9600552283 | -0.2479503836 |
| | C | -0.4571367087 | 0.0115373802 | -3.2238624198 |
| | C | 1.5524508023 | 1.4890979255 | -2.8518281417 |
| | C | -2.0159393139 | 0.9246941880 | -0.3377482764 |
| | C | -2.2834354747 | -0.3509479971 | -0.0496991026 |
| | C | -1.1812343315 | -1.2891203307 | 0.0888161116 |
| | C | -3.6866917326 | -0.8672761375 | 0.1401144469 |
| | H | 1.3846777604 | 0.6900663308 | -0.3597431961 |
| | H | 1.4270606391 | -0.5888336735 | -2.4443582508 |
| | H | -0.5798836275 | 2.2771280132 | -1.1520149567 |
| | H | -0.3371222547 | 1.8755561345 | 0.5141255427 |
| | H | 0.8700396515 | -1.6902880218 | -0.1592491233 |
| | H | -1.1291658999 | 0.8640272239 | -3.2373739503 |

|   |   |   |   |
|---|---|---|---|
| H | -0.1484863682 | -0.1761081315 | -4.2485112680 |
| H | -1.0169182525 | -0.8506241934 | -2.8821676018 |
| H |  0.9691971504 |  2.4031313701 | -2.7936670330 |
| H |  1.8353579867 |  1.3532959175 | -3.8915355551 |
| H |  2.4633077411 |  1.6393811055 | -2.2794116158 |
| H | -2.8180551472 |  1.6365199122 | -0.4391137378 |
| H | -1.3973619549 | -2.2787426053 |  0.4547282112 |
| H | -4.4198229728 | -0.0783248549 |  0.0201542592 |
| H | -3.8116191834 | -1.2949876466 |  1.1318446691 |
| H | -3.9146688072 | -1.6518346421 | -0.5769474763 |

| | |
|---|---|
| $S_0$ energy / H | -387.96001187924026 |
| $S_0$ CI eigenvector | -0.97523499511617  X36 X37 X38 |
| |  0.11531564588785  X36 X37 X39 |
| | -0.09651088379450  X36 X37 A38 B39 |
| | -0.09651088379450  X36 X37 B38 A39 |
| |  0.08816290464405  X36 X38 X39 |
| | -0.05822351679776  X36 A37 B38 X39 |
| | -0.05822351679776  X36 B37 A38 X39 |
| |  0.02938383656857  A36 X37 B38 X39 |
| |  0.02938383656857  B36 X37 A38 X39 |
| $S_1$ energy / H | -387.77070837961162 |
| $S_1$ CI eigenvector | -0.66679503700194  X36 X37 A38 B39 |
| | -0.66679503700194  X36 X37 B38 A39 |
| | -0.16389647648165  X36 A37 X38 B39 |
| | -0.16389647648165  X36 B37 X38 A39 |
| |  0.14315474846456  X36 X37 X39 |
| |  0.13311887925859  X36 X37 X38 |
| | -0.10301841399781  X36 X38 X39 |

0.04697596455500  X36 A37 B38 X39

0.04697596455500  X36 B37 A38 X39

0.03705590102833  A36 X37 B38 X39

0.03705590102833  B36 X37 A38 X39

---

### ax-T S$_1$/S$_0$ MECI-2 (Closed)

| Cartesian coordinates / Å | 26 | | | |
|---|---|---|---|---|
| | C | 0.1808644782 | 0.4500345681 | -1.0668957398 |
| | C | 0.7306901804 | 0.4416699246 | -2.5058899281 |
| | C | -0.9829282960 | 1.4208801042 | -0.7826467665 |
| | C | -0.4381036175 | -0.8472752336 | -0.6259978323 |
| | C | -0.3326682595 | 0.2274610726 | -3.5901252213 |
| | C | 1.5327951603 | 1.7166074448 | -2.7840262730 |
| | C | -2.2117374626 | 0.5646794218 | -0.5562637724 |
| | C | -2.1891123592 | -0.4507695297 | 0.4309142767 |
| | C | -0.9494181186 | -1.0030989701 | 0.7224953486 |
| | C | -3.4824466097 | -1.0271946217 | 0.9467804728 |
| | H | 1.0117094401 | 0.6391196786 | -0.3859767902 |
| | H | 1.4268204076 | -0.3954272541 | -2.5637397676 |
| | H | -1.1399214090 | 2.0923670604 | -1.6169514010 |
| | H | -0.7745655244 | 2.0424358630 | 0.0881733393 |
| | H | -0.4454872489 | -1.6854188197 | -1.3093100023 |
| | H | -1.0214143992 | 1.0636226718 | -3.6468541643 |
| | H | 0.1434552956 | 0.1320520463 | -4.5613210902 |
| | H | -0.9210161497 | -0.6665709663 | -3.4189634304 |
| | H | 0.9088260499 | 2.6009700121 | -2.6971309000 |
| | H | 1.9423490954 | 1.7041018794 | -3.7893597243 |
| | H | 2.3596023127 | 1.8232361887 | -2.0882471015 |

| | | | |
|---|---|---|---|
| H | -3.1628201341 | 0.8978374660 | -0.9351446585 |
| H | -0.8841031802 | -1.9433254484 | 1.2378430318 |
| H | -3.8865408308 | -1.7343607638 | 0.2286516487 |
| H | -4.2189872313 | -0.2438387243 | 1.0937466728 |
| H | -3.3235911130 | -1.5493832365 | 1.8823656843 |

$S_0$ energy / H       -387.80647282293432

$S_0$ CI eigenvector       -0.59113790085849   X36 X37 A38 B39

       -0.59113790085849   X36 X37 B38 A39

       0.51556069399260   X36 X37 X38

       0.11745014710862   X36 A37 B38 X39

       0.11745014710862   X36 B37 A38 X39

       -0.06060959378354   X36 X38 X39

       0.04043277310390   X36 A37 X38 B39

       0.04043277310390   X36 B37 X38 A39

       -0.02238637587029   X37 X38 X39

$S_1$ energy / H       -387.80647265599879

$S_1$ CI eigenvector       -0.83395067122596   X36 X37 X38

       -0.36275405473977   X36 X37 A38 B39

       -0.36275405473977   X36 X37 B38 A39

       -0.10455299633048   X36 A37 X38 B39

       -0.10455299633048   X36 B37 X38 A39

       0.09196027125666   X36 X38 X39

       0.06801797704185   X36 A37 B38 X39

       0.06801797704185   X36 B37 A38 X39

---

**ax-T $S_1/S_0$ MECI-2 (Closed)**

---

| Cartesian coordinates / Å | 26 | | | |
|---|---|---|---|---|
| | C | 0.3825027852 | 0.4227521758 | -0.8695869794 |
| | C | 0.9825591475 | 0.2774215461 | -2.3001824496 |
| | C | -0.7345064912 | 1.5038137890 | -0.7247397624 |
| | C | -0.0544360883 | -0.9383149088 | -0.3781330739 |
| | C | -0.0509064085 | -0.0303514342 | -3.3894865337 |
| | C | 1.8159616934 | 1.5067801491 | -2.6777693529 |
| | C | -2.1972186461 | 0.9933194355 | -0.6416488270 |
| | C | -2.3018615671 | -0.2183401653 | 0.0768021812 |
| | C | -1.2881109985 | -1.2836106789 | -0.0115475886 |
| | C | -3.5024456664 | -0.5108005602 | 0.9254816919 |
| | H | 1.2307323963 | 0.6974915081 | -0.2424082460 |
| | H | 1.6730486977 | -0.5649883173 | -2.2619720432 |
| | H | -0.6608633583 | 2.2147958570 | -1.5360841287 |
| | H | -0.5458803302 | 2.0752356565 | 0.1791964595 |
| | H | 0.7183790077 | -1.6902556425 | -0.3373438151 |
| | H | -0.7414920618 | 0.7943437359 | -3.5271475416 |
| | H | 0.4476300772 | -0.1992868816 | -4.3395103268 |
| | H | -0.6301688817 | -0.9200618565 | -3.1623629051 |
| | H | 1.2018960046 | 2.3943282812 | -2.7865742496 |
| | H | 2.3252935260 | 1.3432384308 | -3.6226938174 |
| | H | 2.5720026887 | 1.7144821443 | -1.9266758971 |
| | H | -2.1563431827 | 0.2968832386 | -1.5502115119 |
| | H | -1.4993820997 | -2.2772528326 | 0.3455303013 |
| | H | -3.8512944056 | 0.4119493517 | 1.3690249772 |
| | H | -3.3176394320 | -1.2741130684 | 1.6746594239 |
| | H | -4.2915500355 | -0.8716910782 | 0.2701318010 |

| $S_0$ energy / H | -387.80219477126423 |
|---|---|
| $S_0$ CI eigenvector | -0.96804494677227  X36 X37 X38 |
|                      | -0.11184870321220  X36 X37 A38 B39               |
|                      | -0.11184870321220  X36 X37 B38 A39               |
|                      |  0.11058939461755  X36 A37 X38 B39               |
|                      |  0.11058939461755  X36 B37 X38 A39               |
|                      |  0.10190087810522  X36 X38 X39                   |
|                      |  0.04912619277670  X37 X38 X39                   |
| S$_1$ energy / H | -387.80219451298774 |
| S$_1$ CI eigenvector | -0.68943496768376  X36 X37 A38 B39 |
| | -0.68943496768376  X36 X37 B38 A39 |
| |  0.15743063927787  X36 X37 X38 |
| | -0.10730489050749  X36 A37 B38 X39 |
| | -0.10730489050749  X36 B37 A38 X39 |

## ax-T S$_1$/S$_0$ MECI-3 (Open)

| | | | | |
|---|---|---|---|---|
| Cartesian coordinates / Å | 26 | | | |
| | C |  0.7600752822 |  0.2619303598 | -0.7345672864 |
| | C |  0.7830153508 |  0.2579166972 | -2.2720168789 |
| | C | -0.7406620023 |  1.6333500825 | -0.0491431933 |
| | C |  0.1707195023 | -0.6765232557 |  0.1320007060 |
| | C | -0.5335351606 |  0.1770755693 | -3.0511408672 |
| | C |  1.6153410954 |  1.4461789347 | -2.7729738947 |
| | C | -1.9938387835 |  1.0912451088 | -0.3986140253 |
| | C | -2.2399739581 | -0.2727143345 | -0.1342096149 |
| | C | -1.1891844003 | -1.1166813565 |  0.1876445025 |
| | C | -3.6503216637 | -0.8011281549 | -0.1663487214 |
| | H |  1.7277392808 |  0.5751582644 | -0.3681904871 |
| | H |  1.3441360304 | -0.6391027403 | -2.5394867067 |

| | | | |
|---|---|---|---|
| H | -0.4100673102 | 2.5368989040 | -0.5340874852 |
| H | -0.3972326430 | 1.5559885046 | 0.9663687377 |
| H | 0.7908265143 | -0.9891812069 | 0.9634273798 |
| H | -1.0920788936 | 1.1032617880 | -3.0062249426 |
| H | -0.3070967046 | -0.0179020666 | -4.0960193106 |
| H | -1.1696708477 | -0.6224615044 | -2.6958546887 |
| H | 1.1271822984 | 2.3925419434 | -2.5577374951 |
| H | 1.7612025856 | 1.3863941928 | -3.8472707733 |
| H | 2.5974036994 | 1.4661655867 | -2.3099577936 |
| H | -2.6080253153 | 1.6098152557 | -1.1132925299 |
| H | -1.3922595925 | -2.0588048690 | 0.6667396840 |
| H | -4.2051156166 | -0.4289940506 | 0.6899114153 |
| H | -3.6767037322 | -1.8840670622 | -0.1553257954 |
| H | -4.1651577275 | -0.4533539167 | -1.0567843432 |

| | |
|---|---|
| $S_0$ energy / H | -387.80336467764687 |
| $S_0$ CI eigenvector | -0.64757822452749  X36 X37 X38 |
| | -0.52600073201432  X36 X37 A38 B39 |
| | -0.52600073201432  X36 X37 B38 A39 |
| | 0.08175196382723  X36 X38 X39 |
| | 0.06172453643308  X37 X38 X39 |
| | 0.05633790510091  A36 X37 B38 X39 |
| | 0.05633790510091  B36 X37 A38 X39 |
| | -0.05592383401504  X36 A37 B38 X39 |
| | -0.05592383401504  X36 B37 A38 X39 |
| | -0.03692629107170  X36 X37 X39 |
| | 0.03170782386289  X36 A37 X38 B39 |
| | 0.03170782386289  X36 B37 X38 A39 |
| $S_1$ energy / H | -387.80336462500168 |

| S₁ CI eigenvector | 0.73768411723328 X36 X37 X38 |
| --- | --- |
| | -0.46159502663600 X36 X37 A38 B39 |
| | -0.46159502663600 X36 X37 B38 A39 |
| | -0.10233865251497 X37 X38 X39 |
| | -0.07053653834627 X36 X38 X39 |
| | 0.04845815828915 X36 X37 X39 |
| | -0.04455288932464 X36 A37 B38 X39 |
| | -0.04455288932464 X36 B37 A38 X39 |
| | 0.03739804278241 A36 X37 X38 B39 |
| | 0.03739804278241 B36 X37 X38 A39 |
| | 0.03634317401199 A36 X37 B38 X39 |
| | 0.03634317401199 B36 X37 A38 X39 |
| | -0.02898498174395 X36 A37 X38 B39 |
| | -0.02898498174395 X36 B37 X38 A39 |

## ax-T cZc-ZDOT Minimum

| Cartesian coordinates / Å | 26 | | |
| --- | --- | --- | --- |
| | C 1.2262137283 | -0.4709174959 | -0.7964103447 |
| | C 1.1585813768 | -0.7202937895 | -2.2886809088 |
| | C -1.3126608512 | 2.1604922434 | -0.0579900102 |
| | C 0.3182307415 | -0.5479889060 | 0.1640824436 |
| | C -0.0561263468 | -1.4957423448 | -2.8020005830 |
| | C 1.3132248648 | 0.6118619887 | -3.0400384177 |
| | C -2.2381946639 | 1.2201001551 | -0.3434485250 |
| | C -2.1974653729 | -0.2160018617 | -0.1001160044 |
| | C -1.1056778271 | -0.9664229034 | 0.1054498748 |
| | C -3.5571628023 | -0.8844251902 | -0.1200457241 |
| | H 2.2107758561 | -0.1605178311 | -0.4785110005 |
| | H 2.0379107697 | -1.3203721912 | -2.5212465027 |
| H | -1.5015201345 | 3.1909582647 | -0.2985648791 |
| H | -0.3728095438 | 1.9375343054 | 0.4052460106 |
| H | 0.6675242169 | -0.3292544311 | 1.1641820086 |
| H | -0.9680672438 | -0.9183182247 | -2.7200228327 |
| H | 0.0896235342 | -1.7428833134 | -3.8498663030 |
| H | -0.1950702827 | -2.4248326024 | -2.2598014966 |
| H | 0.4596576928 | 1.2555880077 | -2.8552361808 |
| H | 1.3879977913 | 0.4428947374 | -4.1104168857 |
| H | 2.2073075204 | 1.1424309579 | -2.7260763152 |
| H | -3.1635580624 | 1.5697249496 | -0.7744022235 |
| H | -1.2690474142 | -2.0115665824 | 0.3179631918 |
| H | -4.2270671192 | -0.4266243520 | 0.6026110694 |
| H | -3.4893275819 | -1.9419200206 | 0.1032179585 |
| H | -4.0217628459 | -0.7781535695 | -1.0974874198 |

| | |
|---|---|
| $S_0$ energy / H | -387.91742606262198 |
| $S_0$ CI eigenvector | -0.97045088079365  X36 X37 X38 |
| | 0.11630963210912  X36 X37 X39 |
| | -0.11088909562897  X36 X37 A38 B39 |
| | -0.11088909562897  X36 X37 B38 A39 |
| | 0.09408824985995  X36 X38 X39 |
| | -0.06885883475258  X36 A37 B38 X39 |
| | -0.06885883475258  X36 B37 A38 X39 |
| $S_1$ energy / H | -387.71958766363451 |
| $S_1$ CI eigenvector | -0.64052538503022  X36 X37 A38 B39 |
| | -0.64052538503022  X36 X37 B38 A39 |
| | -0.21759427849506  X36 A37 X38 B39 |
| | -0.21759427849506  X36 B37 X38 A39 |
| | 0.18894574911363  X36 X37 X39 |

|  |  |  |  |  |
|---|---|---|---|---|
|  |  | 0.14280142485137 | X36 X37 X38 |  |
|  |  | -0.14148612199564 | X36 X38 X39 |  |
|  |  | -0.03989177177915 | A36 X37 X38 B39 |  |
|  |  | -0.03989177177915 | B36 X37 X38 A39 |  |
|  |  | 0.03369976424435 | A36 X37 B38 X39 |  |
|  |  | 0.03369976424435 | B36 X37 A38 X39 |  |
|  |  | -0.03126773855448 | A36 B37 X38 X39 |  |
|  |  | -0.03126773855448 | B36 A37 X38 X39 |  |
|  |  | 0.02478811920746 | X36 A37 B38 X39 |  |
|  |  | 0.02478811920746 | X36 B37 A38 X39 |  |

| **ax-G+** - S$_0$ Minimum (FC) |||||
|---|---|---|---|---|
| Cartesian coordinates / Å | 26 ||||
|  | C |  0.2417848542 |  0.5701366501 | -1.0365749936 |
|  | C |  0.8546216929 |  0.5714765869 | -2.4499952311 |
|  | C | -1.0090832397 |  1.4537513061 | -0.8627792298 |
|  | C | -0.2920112659 | -0.7461200204 | -0.5398810184 |
|  | C |  2.1808000036 | -0.1977077169 | -2.4771258328 |
|  | C | -0.0826431040 |  0.0677752757 | -3.5552569029 |
|  | C | -2.1781085746 |  0.5188223250 | -0.6346404501 |
|  | C | -2.1199382797 | -0.4347732054 |  0.4079666100 |
|  | C | -0.8532907428 | -0.8537565164 |  0.7953521084 |
|  | C | -3.3812709159 | -1.1001204819 |  0.8962881300 |
|  | H |  1.0153605740 |  0.8752799282 | -0.3319253066 |
|  | H |  1.0813164136 |  1.6136136112 | -2.6683074401 |
|  | H | -1.1714313208 |  2.0607451758 | -1.7457945641 |
|  | H | -0.8916230957 |  2.1390167356 | -0.0229087410 |

```
H   -0.2254846849  -1.6212861413  -1.1698089689
H    2.0374899805  -1.2525863255  -2.2564921360
H    2.6464459493  -0.1333569981  -3.4557466112
H    2.8820119421   0.1992920618  -1.7493746399
H   -0.2891439292  -0.9936089832  -3.4589723916
H    0.3777200529   0.2202927156  -4.5269592696
H   -1.0338969041   0.5871290276  -3.5552808382
H   -3.1318636476   0.7526938900  -1.0752606424
H   -0.7413589069  -1.7518304230   1.3744045335
H   -3.6474338275  -1.9240059857   0.2409085387
H   -4.2099822650  -0.3998897440   0.9083994734
H   -3.2353313542  -1.4981302333   1.8930299388
```

| | |
|---|---|
| $S_0$ energy / H | -387.80663156815126 |
| $S_0$ CI eigenvector | -0.58318663210468  X36 X37 A38 B39 |
| | -0.58318663210468  X36 X37 B38 A39 |
| |  0.53343576501733  X36 X37 X38 |
| |  0.11705072203375  X36 A37 B38 X39 |
| |  0.11705072203375  X36 B37 A38 X39 |
| | -0.06118923433112  X36 X38 X39 |
| |  0.04007527920139  X36 A37 X38 B39 |
| |  0.04007527920139  X36 B37 X38 A39 |
| | -0.02315067218226  X37 X38 X39 |
| $S_1$ energy / H | -387.80663149768958 |
| $S_1$ CI eigenvector | -0.82259159908024  X36 X37 X38 |
| | -0.37505867221203  X36 X37 A38 B39 |
| | -0.37505867221203  X36 X37 B38 A39 |
| | -0.10593731455118  X36 A37 X38 B39 |
| | -0.10593731455118  X36 B37 X38 A39 |
| |  0.09019604216599  X36 X38 X39 |

| | | | |
|---|---|---|---|
| 0.06967442377067 | X36 | A37 B38 | X39 |
| 0.06967442377067 | X36 | B37 A38 | X39 |
| 0.03529369718816 | X37 | X38 | X39 |

## Ax-G+ - S$_1$/S$_0$ MECI-1 (Closed)

Cartesian coordinates / Å    26

| | | | |
|---|---|---|---|
| C | 0.1979337252 | 0.7696784185 | -1.1184560652 |
| C | 0.8099382383 | 0.4867837568 | -2.5228037267 |
| C | -1.2188185788 | 1.4210002335 | -1.1895952109 |
| C | 0.2273857230 | -0.4638820890 | -0.2442485828 |
| C | 2.2944929198 | 0.1069655953 | -2.4482692050 |
| C | 0.0329359314 | -0.5417071924 | -3.3524142698 |
| C | -2.4336022205 | 0.5062040787 | -0.8860299446 |
| C | -2.1499743709 | -0.4242483781 | 0.1376037291 |
| C | -0.8353938360 | -1.0739851565 | 0.2811295533 |
| C | -3.2108939813 | -0.8400664046 | 1.1122189883 |
| H | 0.8883631806 | 1.4749939816 | -0.6557015851 |
| H | 0.7561913661 | 1.4380405851 | -3.0464531356 |
| H | -1.3497493330 | 1.8761058939 | -2.1639416020 |
| H | -1.2580989772 | 2.2400051051 | -0.4773086444 |
| H | 1.1998972685 | -0.8773652419 | -0.0332719152 |
| H | 2.4459171497 | -0.8829197728 | -2.0272304194 |
| H | 2.7316716465 | 0.1016806575 | -3.4419085203 |
| H | 2.8562318881 | 0.8158005489 | -1.8464054960 |
| H | -0.0056265028 | -1.5070761617 | -2.8549374287 |
| H | 0.5168976240 | -0.6914094896 | -4.3130089066 |
| H | -0.9835836627 | -0.2211635455 | -3.5469311405 |
| H | -2.1353904445 | -0.3620779845 | -1.5719062675 |
| H | -0.7155419682 | -1.9346002339 | 0.9172099065 |

|   |   |   |   |   |
|---|---|---|---|---|
|   | H | -3.8297808318 | 0.0191453440 | 1.3348112218 |
|   | H | -2.8106144869 | -1.2922825550 | 2.0139165426 |
|   | H | -3.8476339188 | -1.5721954203 | 0.6218562600 |

| S$_0$ energy / H | -387.80181805115842 |
|---|---|
| S$_0$ CI eigenvector | -0.64799097481248  X36 X37 A38 B39 |
|   | -0.64799097481248  X36 X37 B38 A39 |
|   |  0.36632801375322  X36 X37 X38 |
|   | -0.10099624051597  X36 A37 B38 X39 |
|   | -0.10099624051597  X36 B37 A38 X39 |
|   | -0.03812499714040  X36 A37 X38 B39 |
|   | -0.03812499714040  X36 B37 X38 A39 |
|   | -0.03558014074478  X36 X38 X39 |
|   | -0.02337175802583  X37 X38 X39 |

| S$_1$ energy / H | -387.80181795300194 |
|---|---|
| S$_1$ CI eigenvector |  0.90973354915533  X36 X37 X38 |
|   |  0.26066575682312  X36 X37 A38 B39 |
|   |  0.26066575682312  X36 X37 B38 A39 |
|   | -0.10502402148164  X36 A37 X38 B39 |
|   | -0.10502402148164  X36 B37 X38 A39 |
|   | -0.09587225581004  X36 X38 X39 |
|   | -0.04583590102241  X37 X38 X39 |
|   |  0.03757149430946  X36 A37 B38 X39 |
|   |  0.03757149430946  X36 B37 A38 X39 |

**Ax-G+ - S$_1$/S$_0$ MECI-2 (Closed)**

| Cartesian coordinates / Å | 26 |   |   |   |
|---|---|---|---|---|
|   | C | 0.1979337252 | 0.7696784185 | -1.1184560652 |
|   | C | 0.8099382383 | 0.4867837568 | -2.5228037267 |

| | | | |
|---|---|---|---|
| C | -1.2188185788 | 1.4210002335 | -1.1895952109 |
| C | 0.2273857230 | -0.4638820890 | -0.2442485828 |
| C | 2.2944929198 | 0.1069655953 | -2.4482692050 |
| C | 0.0329359314 | -0.5417071924 | -3.3524142698 |
| C | -2.4336022205 | 0.5062040787 | -0.8860299446 |
| C | -2.1499743709 | -0.4242483781 | 0.1376037291 |
| C | -0.8353938360 | -1.0739851565 | 0.2811295533 |
| C | -3.2108939813 | -0.8400664046 | 1.1122189883 |
| H | 0.8883631806 | 1.4749939816 | -0.6557015851 |
| H | 0.7561913661 | 1.4380405851 | -3.0464531356 |
| H | -1.3497493330 | 1.8761058939 | -2.1639416020 |
| H | -1.2580989772 | 2.2400051051 | -0.4773086444 |
| H | 1.1998972685 | -0.8773652419 | -0.0332719152 |
| H | 2.4459171497 | -0.8829197728 | -2.0272304194 |
| H | 2.7316716465 | 0.1016806575 | -3.4419085203 |
| H | 2.8562318881 | 0.8158005489 | -1.8464054960 |
| H | -0.0056265028 | -1.5070761617 | -2.8549374287 |
| H | 0.5168976240 | -0.6914094896 | -4.3130089066 |
| H | -0.9835836627 | -0.2211635455 | -3.5469311405 |
| H | -2.1353904445 | -0.3620779845 | -1.5719062675 |
| H | -0.7155419682 | -1.9346002339 | 0.9172099065 |
| H | -3.8297808318 | 0.0191453440 | 1.3348112218 |
| H | -2.8106144869 | -1.2922825550 | 2.0139165426 |
| H | -3.8476339188 | -1.5721954203 | 0.6218562600 |

| | |
|---|---|
| $S_0$ energy / H | -387.80181805115842 |
| $S_0$ CI eigenvector | -0.64799097481248  X36 X37 A38 B39 |
| | -0.64799097481248  X36 X37 B38 A39 |
| | 0.36632801375322  X36 X37 X38 |
| | -0.10099624051597  X36 A37 B38 X39 |
|  | -0.10099624051597 X36 B37 A38 X39 |
|  | -0.03812499714040 X36 A37 X38 B39 |
|  | -0.03812499714040 X36 B37 X38 A39 |
|  | -0.03558014074478 X36 X38 X39 |
|  | -0.02337175802583 X37 X38 X39 |
| S$_1$ energy / H | -387.80181795300194 |
| S$_1$ CI eigenvector | 0.90973354915533 X36 X37 X38 |
|  | 0.26066575682312 X36 X37 A38 B39 |
|  | 0.26066575682312 X36 X37 B38 A39 |
|  | -0.10502402148164 X36 A37 X38 B39 |
|  | -0.10502402148164 X36 B37 X38 A39 |
|  | -0.09587225581004 X36 X38 X39 |
|  | -0.04583590102241 X37 X38 X39 |
|  | 0.03757149430946 X36 A37 B38 X39 |
|  | 0.03757149430946 X36 B37 A38 X3 |

**Ax-G+ - S$_1$/S$_0$ MECI-3 (Open)**

| | | | | |
|---|---|---|---|---|
| Cartesian coordinates / Å | 26 | | | |
| | C | 0.7287161387 | 0.2030187308 | -0.8467286336 |
| | C | 0.6470982265 | 0.2015435846 | -2.3786516980 |
| | C | -0.6802286569 | 1.5984392577 | -0.0854244165 |
| | C | 0.1821062931 | -0.7173309423 | 0.0597495139 |
| | C | 1.7227340677 | -0.7705250334 | -2.8997429549 |
| | C | -0.6822165347 | -0.1306513745 | -3.0620960643 |
| | C | -1.9711535468 | 1.0945290671 | -0.3590505622 |
| | C | -2.2255840353 | -0.2713006541 | -0.1132311535 |
| | C | -1.1823834620 | -1.1409658490 | 0.1586514114 |
| | C | -3.6416280126 | -0.7829545171 | -0.1295145312 |
| | H | 1.7133214126 | 0.5342864678 | -0.5464606234 |
| | H | 0.9366131510 | 1.2003190121 | -2.7016356517 |

|   |                |               |               |
|---|---------------:|--------------:|--------------:|
| H | -0.3452279821  |  2.4832470546 | -0.6035924735 |
| H | -0.2859495416  |  1.5302793727 |  0.9130282360 |
| H |  0.8322570126  | -1.0169782959 |  0.8729685086 |
| H |  1.4641387215  | -1.7927249911 | -2.6375792745 |
| H |  1.8026831650  | -0.7110449861 | -3.9818652319 |
| H |  2.7006528636  | -0.5566226713 | -2.4802207014 |
| H | -1.0652241982  | -1.0945575777 | -2.7496025960 |
| H | -0.5194412703  | -0.1702249153 | -4.1356342580 |
| H | -1.4398831915  |  0.6138642010 | -2.8720042665 |
| H | -2.6104205837  |  1.6355815820 | -1.0341993391 |
| H | -1.3860072978  | -2.0946448203 |  0.6136039693 |
| H | -4.1764615510  | -0.4397433846 |  0.7508253112 |
| H | -3.6751287676  | -1.8651913987 | -0.1608918579 |
| H | -4.1693448474  | -0.3969120278 | -0.9961449871 |

| | |
|---|---|
| S$_0$ energy / H | -387.80483862897137 |
| S$_0$ CI eigenvector | -0.95975759923633  X36 X37 X38 |
| | -0.14781598943905  X36 X37 A38 B39 |
| | -0.14781598943905  X36 X37 B38 A39 |
| |  0.11114463245535  X37 X38 X39 |
| |  0.10508230583397  X36 X38 X39 |
| | -0.05190930957953  X36 X37 X39 |
| |  0.04650013862945  X36 A37 X38 B39 |
| |  0.04650013862945  X36 B37 X38 A39 |
| | -0.02803180795458  A36 B37 X38 X39 |
| | -0.02803180795458  B36 A37 X38 X39 |
| | -0.02531857949088  A36 X37 X38 B39 |
| | -0.02531857949088  B36 X37 X38 A39 |
| |  0.02438265916908  A36 X37 B38 X39 |
| |  0.02438265916908  B36 X37 A38 X39 |

| | | |
|---|---|---|
| S₁ energy / H | -387.80483844324084 | |
| S₁ CI eigenvector | -0.68375200325077 | X36 X37 A38 B39 |
| | -0.68375200325077 | X36 X37 B38 A39 |
| |  0.20754459505639 | X36 X37 X38 |
| | -0.07332504654601 | X36 A37 B38 X39 |
| | -0.07332504654601 | X36 B37 A38 X39 |
| |  0.06002297019828 | A36 X37 B38 X39 |
| |  0.06002297019828 | B36 X37 A38 X39 |
| | -0.04455957191723 | X37 X38 X39 |
| |  0.02801839030162 | A36 X37 X38 B39 |
| |  0.02801839030162 | B36 X37 X38 A3 |

## Ax-G+ cZc-ZDOT Minimum

| | | | | |
|---|---|---|---|---|
| Cartesian coordinates / Å | 26 | | | |
| | C |  1.2167540946 | -0.4263481655 | -0.8569819289 |
| | C |  1.1402182129 | -0.6871393715 | -2.3472869257 |
| | C | -1.3497894335 |  2.1558913897 | -0.0846773421 |
| | C |  0.3260912437 | -0.5193955953 |  0.1185604699 |
| | C |  2.3580028781 | -1.5247600642 | -2.7688914272 |
| | C | -0.1492815735 | -1.3292490863 | -2.8614041022 |
| | C | -2.2875347749 |  1.2022024154 | -0.2677975836 |
| | C | -2.2046778391 | -0.2343365681 | -0.0422410692 |
| | C | -1.0901570403 | -0.9671671327 |  0.0976530589 |
| | C | -3.5501090893 | -0.9300777802 |  0.0010500629 |
| | H |  2.2000225813 | -0.1019978867 | -0.5497176457 |
| | H |  1.2362961754 |  0.2857772751 | -2.8299142516 |
| | H | -1.5838347448 |  3.1837912708 | -0.2947945179 |
| | H | -0.3609873960 |  1.9491448650 |  0.2707039006 |

```
H    0.6874805475  -0.2778796620   1.1091982699
H    2.3282822282  -2.5077366650  -2.3071481271
H    2.3786983156  -1.6607554925  -3.8456788845
H    3.2893071565  -1.0485454934  -2.4773758954
H   -0.3328087981  -2.2867461875  -2.3840971873
H   -0.0679507446  -1.5036173543  -3.9304102188
H   -1.0101961004  -0.6975264821  -2.6904588913
H   -3.2579232981   1.5403941351  -0.5961475093
H   -1.2230583946  -2.0206418042   0.2881621825
H   -4.1769896962  -0.5126956589   0.7845108326
H   -3.4486583050  -1.9931495903   0.1804553947
H   -4.0825162054  -0.7973653108  -0.9376006641
```

| | | |
|---|---|---|
| $S_0$ energy / H | -387.91686946030802 | |
| $S_0$ CI eigenvector | -0.96934101487212 | X36 X37 X38 |
| | 0.11972552996098 | X36 X37 X39 |
| | -0.11476083098522 | X36 X37 A38 B39 |
| | -0.11476083098522 | X36 X37 B38 A39 |
| | 0.09461400324175 | X36 X38 X39 |
| | 0.06827796568446 | X36 A37 B38 X39 |
| | 0.06827796568446 | X36 B37 A38 X39 |
| | | |
| $S_1$ energy / H | -387.7188817407411 | |
| $S_1$ CI eigenvector | 0.63712461440479 | X36 X37 A38 B39 |
| | 0.63712461440479 | X36 X37 B38 A39 |
| | -0.22171951676535 | X36 A37 X38 B39 |
| | -0.22171951676535 | X36 B37 X38 A39 |
| | -0.19783083056383 | X36 X37 X39 |
| | -0.14841740420460 | X36 X37 X38 |
| | 0.14788885665153 | X36 X38 X39 |

0.03408096242917  A36 X37 X38 B39

0.03408096242917  B36 X37 X38 A39

-0.02928668817313  A36 X37 B38 X39

-0.02928668817313  B36 X37 A38 X39

-0.02661533151651  A36 B37 X38 X39

-0.02661533151651  B36 A37 X38 X39

0.02579770266313  X36 A37 B38 X39

0.02579770266313  X36 B37 A38 X39

---

### eq-G- S$_0$ Minimum (FC)

| Cartesian coordinates / Å | 26 | | | |
|---|---|---|---|---|
| | C | 0.7215842185 | -0.0176476731 | 0.1956837804 |
| | C | 2.1871475914 | -0.0476244843 | -0.2973649979 |
| | C | -0.0231118548 | 1.2737258499 | -0.1832592245 |
| | C | -0.0785385643 | -1.2088974162 | -0.2865091147 |
| | C | 2.9437810903 | 1.2482435975 | 0.0212230012 |
| | C | 2.9593750992 | -1.2324084311 | 0.2987129588 |
| | C | -1.5046217313 | 1.1881932906 | 0.0822919164 |
| | C | -2.1705195596 | 0.0346739532 | -0.0151504071 |
| | C | -1.4270730878 | -1.1696272090 | -0.3448711194 |
| | C | -3.6615937916 | -0.0703838162 | 0.1758983077 |
| | H | 0.7459488470 | -0.0627673021 | 1.2883797787 |
| | H | 2.1685409125 | -0.1643166350 | -1.3802441370 |
| | H | 0.1314014948 | 1.4797909357 | -1.2453625462 |
| | H | 0.3920133693 | 2.1165617483 | 0.3563176005 |
| | H | 0.4328696946 | -2.1208024234 | -0.5332548962 |
| | H | 2.9163864131 | 1.4643798018 | 1.0867630772 |
| | H | 3.9864562843 | 1.1578140542 | -0.2675142581 |
| H | 2.5366999143 | 2.1042667572 | -0.5037653823 | |
| H | 3.0311326718 | -1.1384796418 | 1.3793814673 | |
| H | 3.9700914053 | -1.2659107955 | -0.0963739498 | |
| H | 2.4969600581 | -2.1882085833 | 0.0825018220 | |
| H | -2.0292503802 | 2.1024847742 | 0.3029085326 | |
| H | -1.9777469460 | -2.0500451328 | -0.6305709975 | |
| H | -4.1021050295 | 0.8942966314 | 0.3986780990 | |
| H | -3.9038770024 | -0.7495828821 | 0.9891074074 | |
| H | -4.1393511169 | -0.4605289679 | -0.719406718 | |

$S_0$ energy / H

$S_0$ CI eigenvector        -387.96100444601700

    -0.97519068229222   X36 X37 X38

    0.11632710709177   X36 X37 X39

    0.09633384418636   X36 X37 A38 B39

    0.09633384418636   X36 X37 B38 A39

    0.08682545112113   X36 X38 X39

    -0.05806737301177   X36 A37 B38 X39

    -0.05806737301177   X36 B37 A38 X39

    0.03061420004001   A36 X37 B38 X39

    0.03061420004001   B36 X37 A38 X39

$S_1$ energy / H

$S_1$ CI eigenvector        -387.77243518781489

    -0.66680452344343   X36 X37 A38 B39

    -0.66680452344343   X36 X37 B38 A39

    -0.16384811375325   X36 A37 X38 B39

    -0.16384811375325   X36 B37 X38 A39

    -0.14076336339304   X36 X37 X39

    -0.13290784952594   X36 X37 X38

    0.09926063151556   X36 X38 X39

    -0.04867673559357   X36 A37 B38 X39

| | | | |
|---|---|---|---|
| -0.04867673559357 | X36 | B37 A38 | X39 |
| -0.03474854513125 | A36 | X37 B38 | X39 |
| -0.03474854513125 | B36 | X37 A38 | X39 |
| 0.03145485319226 | A36 | X37 X38 | B39 |
| 0.03145485319226 | B36 | X37 X38 | A39 |

## eq-G- S$_1$/S$_0$ MECI-1 (Closed)

Cartesian coordinates / Å    26

| | | | |
|---|---|---|---|
| C | 0.8205846739 | 0.2056237093 | 0.5701030249 |
| C | 2.0239412939 | 0.0476758007 | -0.3772567085 |
| C | -0.0491291391 | 1.4578870690 | 0.3775027571 |
| C | -0.2250101521 | -0.8674528447 | 0.3842313533 |
| C | 3.0040730779 | 1.2122506109 | -0.2100399871 |
| C | 2.7322354646 | -1.2939719954 | -0.1578851361 |
| C | -1.4760188018 | 1.0100455143 | 0.6203765233 |
| C | -2.0257626551 | 0.0051221609 | -0.2086215388 |
| C | -1.1090933929 | -0.8737839235 | -0.7721684666 |
| C | -3.5166093311 | -0.2054856091 | -0.2836208056 |
| H | 1.1706646381 | 0.1724767687 | 1.5994590956 |
| H | 1.6364802501 | 0.0646366178 | -1.3919447569 |
| H | 0.0476778765 | 1.8443099289 | -0.6371753129 |
| H | 0.2377228012 | 2.2495221910 | 1.0592224010 |
| H | -0.2962895087 | -1.6526282735 | 1.1242266122 |
| H | 3.4024449218 | 1.2467612720 | 0.8011088545 |
| H | 3.8430192727 | 1.1104875417 | -0.8919962212 |
| H | 2.5316097121 | 2.1673471756 | -0.4132297416 |
| H | 3.0949339780 | -1.3856679017 | 0.8633568841 |
| H | 3.5861915268 | -1.3958827963 | -0.8205842000 |
| H | 2.0701805684 | -2.1329565228 | -0.3525684046 |
| H | -2.1230498577 | 1.5944887338 | 1.2504867304 |
| H | -1.4321210927 | -1.7893763946 | -1.2317389901 |
| H | -4.0564177275 | 0.7185871082 | -0.1070306019 |
| H | -3.8322956727 | -0.9320161742 | 0.4593472024 |
| H | -3.7848547101 | -0.5927407573 | -1.2599603127 |

| | |
|---|---|
| S$_0$ energy / H | -387.80956229542915 |
| S$_0$ CI eigenvector | 0.58512611242490  X36 X37 X38 |
| | -0.55711001454199  X36 X37 A38 B39 |
| | -0.55711001454199  X36 X37 B38 A39 |
| | 0.11460965090409  X36 A37 B38 X39 |
| | 0.11460965090409  X36 B37 A38 X39 |
| | -0.06988733178561  X36 X38 X39 |
| | 0.04865608199524  X36 A37 X38 B39 |
| | 0.04865608199524  X36 B37 X38 A39 |
| | -0.02509910410607  X37 X38 X39 |
| S$_1$ energy / H | -387.80956226639279 |
| S$_1$ CI eigenvector | -0.78540814846225  X36 X37 X38 |
| | -0.41198449833168  X36 X37 A38 B39 |
| | -0.41198449833168  X36 X37 B38 A39 |
| | -0.10461145972005  X36 A37 X38 B39 |
| | -0.10461145972005  X36 B37 X38 A39 |
| | 0.08889472040544  X36 X38 X39 |
| | 0.07822105820126  X36 A37 B38 X39 |
| | 0.07822105820126  X36 B37 A38 X39 |
| | 0.03351048927819  X37 X38 X39 |

**eq-G- S$_1$/S$_0$ MECI-2 (Closed)**

Cartesian coordinates / Å

| | | | | |
|---|---|---|---|---|
| C | 0.7724558188 | 0.0881489672 | 0.2575458415 |
| C | 2.1641725098 | -0.0301612690 | -0.4149372244 |
| C | -0.0953169535 | 1.2859829156 | -0.2199619392 |
| C | 0.0013187188 | -1.2022528236 | 0.1664981594 |
| C | 2.8643924751 | 1.3301594762 | -0.5178322918 |
| C | 3.0899203241 | -1.0125060859 | 0.3162115250 |
| C | -1.5893856116 | 1.2244938718 | 0.2010323763 |
| C | -2.1032783329 | -0.0733967160 | -0.0479152003 |
| C | -1.3347114484 | -1.2994743567 | 0.1253622107 |
| C | -3.5198039553 | -0.2291249812 | -0.5082117168 |
| H | 0.9334500337 | 0.2393880853 | 1.3310635797 |
| H | 2.0086792564 | -0.3982033194 | -1.4288842151 |
| H | -0.0528859469 | 1.3526859865 | -1.3066551869 |
| H | 0.3385638732 | 2.2015648128 | 0.1573238533 |
| H | 0.5648228322 | -2.1195474218 | 0.1592786458 |
| H | 2.3133401099 | 2.0305406558 | -1.1321480463 |
| H | 2.9925329792 | 1.7761990325 | 0.4652017450 |
| H | 3.8500695782 | 1.2125511006 | -0.9572133342 |
| H | 3.2938458834 | -0.6673817568 | 1.3264566539 |
| H | 4.0407637876 | -1.0957321960 | -0.2006340527 |
| H | 2.6799381386 | -2.0142489723 | 0.3880038954 |
| H | -1.4044340302 | 0.8469200358 | 1.2705441974 |
| H | -1.8086803251 | -2.2636121218 | 0.0569376242 |
| H | -4.1630174545 | 0.0337230138 | 0.3259636754 |
| H | -3.7733680767 | -1.2266639856 | -0.8496290174 |
| H | -3.6979371855 | 0.5169857453 | -1.2733276221 |

| | | |
|---|---|---|
| S$_0$ energy / H | -387.80592223014224 | |
| S$_0$ CI eigenvector | 0.69616470434947 | X36 X37 A38 B39 |
| | 0.69616470434947 | X36 X37 B38 A39 |
| | 0.09227620068639 | X36 X37 X38 |
| | -0.08860596014356 | A36 X37 B38 X39 |
| | -0.08860596014356 | B36 X37 A38 X39 |
| | -0.05353036383453 | X36 A37 B38 X39 |
| | -0.05353036383453 | X36 B37 A38 X39 |
| | | |
| S$_1$ energy / H | -387.80592209503504 | |
| S$_1$ CI eigenvector | -0.97776450793694 | X36 X37 X38 |
| | 0.08651795125359 | X37 X38 X39 |
| | -0.08511072749567 | X36 A37 X38 B39 |
| | -0.08511072749567 | X36 B37 X38 A39 |
| | 0.06703208733617 | X36 X38 X39 |
| | 0.06576866316932 | X36 X37 A38 B39 |
| | 0.06576866316932 | X36 X37 B38 A39 |
| | -0.06375019652245 | A36 X37 X38 B39 |
| | -0.06375019652245 | B36 X37 X38 A39 |

**eq-G- S$_1$/S$_0$ MECI-3 (Open)**

| | | | | |
|---|---|---|---|---|
| Cartesian coordinates / Å | 26 | | | |
| | C | 0.8645538360 | -0.0952671421 | 0.0089172085 |
| | C | 2.3611565855 | -0.2204773165 | -0.2170248598 |
| | C | -0.4798137648 | 1.6344741216 | -0.6225996939 |
| | C | -0.0263071444 | -0.9367252573 | -0.6613854982 |
| | C | 3.0818100555 | 1.0908440900 | 0.1151330204 |
| | C | 2.9729992878 | -1.3780238197 | 0.5920812997 |
| | C | -1.6489315323 | 1.3311922368 | 0.1035809977 |

```
C   -2.1940256412    0.0547380452   -0.0246396753
C   -1.4364637360   -0.9692238870   -0.6222635145
C   -3.5818141064   -0.2367650660    0.4785190911
H    0.5757732206    0.2918034307    0.9704747946
H    2.5348859884   -0.4326971590   -1.2710366892
H   -0.4065519467    1.3349330604   -1.6502217991
H    0.1155975129    2.4877297847   -0.3504589850
H    0.4226522944   -1.6489164584   -1.3425565291
H    2.8964228245    1.3804213742    1.1467215403
H    4.1564963508    0.9939998763   -0.0104347429
H    2.7429862726    1.9010520633   -0.5229560802
H    2.8173893362   -1.2293701922    1.6572890695
H    4.0439384611   -1.4575138773    0.4188288234
H    2.5205059167   -2.3274785958    0.3227543418
H   -2.0049691761    2.0113206899    0.8566727426
H   -1.9360179036   -1.8325148995   -1.0244789700
H   -4.0772366140    0.6582876061    0.8338679088
H   -3.5503935388   -0.9690005425    1.2793224344
H   -4.1784777464   -0.6631974590   -0.3227357354
```

| | | |
|---|---|---|
| $S_0$ energy / H | -387.82385202031855 | |
| $S_0$ CI eigenvector | -0.69350859306081 | X36 X37 A38 B39 |
| | -0.69350859306081 | X36 X37 B38 A39 |
| |  0.13019679049011 | X36 X37 X38 |
| | -0.09760916431948 | X36 A37 B38 X39 |
| | -0.09760916431948 | X36 B37 A38 X39 |
| | -0.03104039399901 | X36 X38 X39 |
| |  0.02315939544216 | X36 X37 X3 |

| | |
|---|---|
| $S_1$ energy / H | -387.82385189660880 |

| S$_1$ CI eigenvector | 0.97107198284765 | X36 X37 X38 |
| --- | --- | --- |
| | -0.14405395540441 | X36 X38 X39 |
| | 0.09587170452449 | X36 X37 A38 B39 |
| | 0.09587170452449 | X36 X37 B38 A39 |
| | -0.08012434328866 | X37 X38 X39 |
| | 0.06549146721486 | X36 X37 X39 |
| | -0.05782472008042 | X36 A37 X38 B39 |
| | -0.05782472008042 | X36 B37 X38 A39 |

## eq-G- cZc-EDOT Minimum

| Cartesian coordinates / Å | 26 | | |
| --- | --- | --- | --- |
| | C   1.1304329602 | -0.6808460378 |  0.1531111476 |
| | C   2.6108857186 | -0.8726431806 | -0.0582562878 |
| | C  -0.8397391593 |  2.1781603942 | -0.3330961895 |
| | C   0.1826872384 | -1.1238807229 | -0.6662253754 |
| | C   3.3168251808 |  0.4797027822 | -0.2296180574 |
| | C   3.2280019423 | -1.6642512586 |  1.1027154228 |
| | C  -1.5180011267 |  1.3362282260 |  0.4263986706 |
| | C  -2.0141919731 |  0.0035058644 |  0.0151788084 |
| | C  -1.2671213438 | -0.9977242344 | -0.5134790172 |
| | C  -3.5047389718 | -0.1731498095 |  0.2036390696 |
| | H   0.8513437450 | -0.1391939872 |  1.0426050131 |
| | H   2.7529239070 | -1.4449278437 | -0.9718693548 |
| | H  -0.5348543828 |  1.9197469593 | -1.3310549044 |
| | H  -0.5581763229 |  3.1524309024 |  0.0257170257 |
| | H   0.4921310612 | -1.7089542178 | -1.5203398084 |
| | H   3.1929135743 |  1.0984489536 |  0.6553718593 |
| | H   4.3822420307 |  0.3430806567 | -0.3911941615 |
| | H   2.9144137370 |  1.0254576344 | -1.0767109078 |
| H | 3.1046789773 | -1.1340957527 | 2.0435838755 |
| H | 4.2918904388 | -1.8178571709 | 0.9460700741 |
| H | 2.7577650806 | -2.6366969437 | 1.2048889576 |
| H | -1.8250750906 | 1.6676335611 | 1.4084721413 |
| H | -1.8128539633 | -1.8591925958 | -0.8652825050 |
| H | -3.7805972017 | -0.0713139520 | 1.2518827207 |
| H | -3.8350880037 | -1.1493028144 | -0.1308188385 |
| H | -4.0639280527 | 0.5824345881 | -0.3433293788 |

| | |
|---|---|
| $S_0$ energy / H | -387.92638919143826 |
| $S_0$ CI eigenvector | 0.98729585419654  X36 X37 X38 |
| | -0.09333196538926  X36 X37 X39 |
| | -0.07379587045791  X37 X38 X39 |
| | -0.05689852958353  A36 X37 B38 X39 |
| | -0.05689852958353  B36 X37 A38 X39 |
| | 0.03806106086014  X36 X37 A38 B39 |
| | 0.03806106086014  X36 X37 B38 A39 |
| | -0.03748386015648  X36 X38 X39 |
| $S_1$ energy / H | -387.73789017951913 |
| $S_1$ CI eigenvector | 0.68854109167997  X36 X37 A38 B39 |
| | 0.68854109167997  X36 X37 B38 A39 |
| | -0.11757409596283  A36 X37 X38 B39 |
| | -0.11757409596283  B36 X37 X38 A39 |
| | -0.09020469298112  X36 A37 B38 X39 |
| | -0.09020469298112  X36 B37 A38 X39 |
| | -0.05276435889037  X36 X37 X38 |
| | -0.05123689289574  X36 X37 X39 |
| | 0.03517039062917  X37 X38 X39 |
| | 0.02328033685207  X36 A37 X38 B39 |

0.02328033685207  X36 B37 X38 A39

| | **eq-T S$_0$ Minimum (FC)** | | | |
|---|---|---|---|---|
| Cartesian coordinates / Å | 26 | | | |
| | C | 0.7057597469 | 0.0237544854 | 0.1411151022 |
| | C | 2.1723001798 | -0.0213651958 | -0.3431453567 |
| | C | -0.1260218952 | 1.1862578378 | -0.4248841664 |
| | C | -0.0132019614 | -1.2857727150 | -0.0859510646 |
| | C | 2.3222236753 | -0.2396929353 | -1.8536175369 |
| | C | 2.9686119337 | 1.2090630395 | 0.1040870397 |
| | C | -1.5944317799 | 1.0447811606 | -0.1112177847 |
| | C | -2.1788127838 | -0.1508849697 | 0.0088839364 |
| | C | -1.3617565860 | -1.3479548047 | -0.1079293012 |
| | C | -3.6580792044 | -0.3159238124 | 0.2468169455 |
| | H | 0.7446689661 | 0.1675857197 | 1.2242459065 |
| | H | 2.6172269885 | -0.8844624529 | 0.1506684005 |
| | H | -0.0140835236 | 1.2329888968 | -1.5085834025 |
| | H | 0.2422429672 | 2.1306146352 | -0.0390170489 |
| | H | 0.5697536686 | -2.1895358151 | -0.1363372372 |
| | H | 1.9850938788 | 0.6221323403 | -2.4208210383 |
| | H | 3.3653477121 | -0.4044559118 | -2.1072800077 |
| | H | 1.7583117501 | -1.1011605270 | -2.1955237782 |
| | H | 2.6314201107 | 2.1106839297 | -0.3982812198 |
| | H | 4.0228904957 | 1.0862604543 | -0.1254590171 |
| | H | 2.8800568115 | 1.3709888251 | 1.1745859597 |
| | H | -2.1823613767 | 1.9447068789 | -0.0466656314 |
| | H | -1.8567961662 | -2.3008172372 | -0.1900486800 |
| | H | -4.1601928511 | 0.6425655125 | 0.3052079878 |
| | H | -3.8468681307 | -0.8537281379 | 1.1723299508 |

H    -4.1173026260  -0.8886292005  -0.5551789576

| | |
|---|---|
| $S_0$ energy / H | -387.96159429897074 |
| $S_0$ CI eigenvector | -0.97493281468196  X36 X37 X38 |
| |  0.11662336390680  X36 X37 X39 |
| | -0.09735180221812  X36 X37 A38 B39 |
| | -0.09735180221812  X36 X37 B38 A39 |
| |  0.08655202046549  X36 X38 X39 |
| | -0.05799785652372  X36 A37 B38 X39 |
| | -0.05799785652372  X36 B37 A38 X39 |
| |  0.03116916936685  A36 X37 B38 X39 |
| |  0.03116916936685  B36 X37 A38 X39 |
| $S_1$ energy / H | -387.77315039506851 |
| $S_1$ CI eigenvector | -0.66640928234247  X36 X37 A38 B39 |
| | -0.66640928234247  X36 X37 B38 A39 |
| | -0.16438042434047  X36 A37 X38 B39 |
| | -0.16438042434047  X36 B37 X38 A39 |
| |  0.14207072500754  X36 X37 X39 |
| |  0.13453150546971  X36 X37 X38 |
| | -0.09879323528164  X36 X38 X39 |
| |  0.04850522885015  X36 A37 B38 X39 |
| |  0.04850522885015  X36 B37 A38 X39 |
| |  0.03455292093703  A36 X37 B38 X39 |
| |  0.03455292093703  B36 X37 A38 X39 |
| |  0.03128328492401  A36 X37 X38 B39 |
| |  0.03128328492401  B36 X37 X38 A39 |

**eq-T $S_1$/$S_0$ MECI-1 (Closed)**

| | | | | |
|---|---|---|---|---|
| Cartesian coordinates / Å | 26 | | | |
| | C | 0.7607767557 | 0.2409503832 | 0.4054968458 |
| | C | 2.0450281606 | 0.0504225558 | -0.4355621621 |
| | C | -0.1295918459 | 1.4669736053 | 0.1750342942 |
| | C | -0.2731490417 | -0.8573214356 | 0.2861859717 |
| | C | 1.8248738289 | -0.1023877659 | -1.9450035070 |
| | C | 3.0280385578 | 1.1923050715 | -0.1487110497 |
| | C | -1.5103882391 | 1.0098828195 | 0.5972806452 |
| | C | -2.1354034838 | -0.0274553305 | -0.1450278911 |
| | C | -1.2713280193 | -0.9073077576 | -0.7732287345 |
| | C | -3.6219740359 | -0.2476692916 | -0.0459023764 |
| | H | 1.0697267814 | 0.2605708474 | 1.4475878802 |
| | H | 2.5075801987 | -0.8706427150 | -0.0804400936 |
| | H | -0.1415200431 | 1.7610378429 | -0.8734459666 |
| | H | 0.2026411342 | 2.3188512508 | 0.7556178151 |
| | H | -0.2311649265 | -1.6637149295 | 1.0069039108 |
| | H | 1.4002192874 | 0.7941514475 | -2.3834803788 |
| | H | 2.7769793192 | -0.2892192569 | -2.4347360761 |
| | H | 1.1585288062 | -0.9224551794 | -2.1769922009 |
| | H | 2.6359064585 | 2.1441293527 | -0.4952392475 |
| | H | 3.9721958077 | 1.0227971808 | -0.6575899357 |
| | H | 3.2359829075 | 1.2846150026 | 0.9134345938 |
| | H | -2.0917254058 | 1.5943327642 | 1.2881794798 |
| | H | -1.6186748439 | -1.8383577149 | -1.1799877587 |
| | H | -4.1467999680 | 0.6964766869 | 0.0560510525 |
| | H | -3.8486061015 | -0.8545250142 | 0.8257667908 |
| | H | -3.9883817482 | -0.7698450793 | -0.9217480950 |

| | |
|---|---|
| S$_0$ energy / H | -387.80500974883802 |
| S$_0$ CI eigenvector | 0.97192763752711  X36 X37 X38 |
|  | -0.11745472841741  X36 X38 X39 |
|  | 0.11160470710375  X36 A37 X38 B39 |
|  | 0.11160470710375  X36 B37 X38 A39 |
|  | -0.08250169116511  X36 X37 A38 B39 |
|  | -0.08250169116511  X36 X37 B38 A39 |
|  | -0.04076355267413  X37 X38 X39 |
|  | -0.02371222048129  X36 X37 X39 |
| $S_1$ energy / H | -387.80500962784629 |
| $S_1$ CI eigenvector | 0.68762268207827  X36 X37 A38 B39 |
|  | 0.68762268207827  X36 X37 B38 A39 |
|  | -0.13985120101163  X36 A37 B38 X39 |
|  | -0.13985120101163  X36 B37 A38 X39 |
|  | 0.11400361728325  X36 X37 X38 |
|  | 0.02961637043922  X36 A37 X38 B39 |
|  | 0.02961637043922  X36 B37 X38 A39 |

## eq-T $S_1/S_0$ MECI-2 (Closed)

| Cartesian coordinates / Å | 26 | | | |
|---|---|---|---|---|
|  | C | 0.7687321046 | 0.1112457476 | 0.2414067123 |
|  | C | 2.1687557267 | -0.0107788136 | -0.4069880475 |
|  | C | -0.1656619577 | 1.2190177012 | -0.3188610698 |
|  | C | 0.0825120003 | -1.2247692879 | 0.3026601893 |
|  | C | 2.1141315226 | -0.4418339439 | -1.8783183976 |
|  | C | 2.9729331498 | 1.2830718783 | -0.2526130910 |
|  | C | -1.6540457600 | 1.0944556487 | 0.1131239454 |
|  | C | -2.0843954276 | -0.2534453343 | 0.0020587929 |
|  | C | -1.2441465497 | -1.4075437311 | 0.2982265965 |
|  | C | -3.4898173244 | -0.5468893562 | -0.4238838795 |

|   |                |               |               |
|---|----------------|---------------|---------------|
| H |  0.9296824607  |  0.3662585649 |  1.2947943040 |
| H |  2.7022737959  | -0.7884113381 |  0.1406966856 |
| H | -0.1334649623  |  1.2237338250 | -1.4053593580 |
| H |  0.2128612604  |  2.1813995400 | -0.0011049598 |
| H |  0.7176552703  | -2.0920771836 |  0.3975985709 |
| H |  1.5724124368  | -1.3745768407 | -2.0098686396 |
| H |  1.6313701980  |  0.3074606847 | -2.4959843021 |
| H |  3.1163741752  | -0.5900020278 | -2.2686048355 |
| H |  2.5428846835  |  2.0901155110 | -0.8358716046 |
| H |  3.9942066581  |  1.1414730802 | -0.5925145718 |
| H |  3.0107000051  |  1.6047814284 |  0.7840649706 |
| H | -1.4455860259  |  0.8419909603 |  1.2144988436 |
| H | -1.6622656444  | -2.3977957067 |  0.3508660194 |
| H | -4.1465833977  | -0.2225633692 |  0.3767652152 |
| H | -3.6833151334  | -1.5921435157 | -0.6391440536 |
| H | -3.7124993752  |  0.0873558888 | -1.2737004909 |

| | | |
|---|---|---|
| $S_0$ energy / H | -387.80720210961749 | |
| $S_0$ CI eigenvector | -0.63870299452758 | X36 X37 A38 B39 |
| | -0.63870299452758 | X36 X37 B38 A39 |
| |  0.39972856902206 | X36 X37 X38 |
| |  0.07399792765471 | A36 X37 B38 X39 |
| |  0.07399792765471 | B36 X37 A38 X39 |
| | -0.06172236205932 | X36 A37 B38 X39 |
| | -0.06172236205932 | X36 B37 A38 X39 |
| | -0.04076503536987 | X36 A37 X38 B39 |
| | -0.04076503536987 | X36 B37 X38 A39 |
| | -0.03600000405787 | X37 X38 X39 |
| | -0.02968698031213 | X36 X38 X3 |
| $S_1$ energy / H | -387.80720201363135 | |

S$_1$ CI eigenvector

| | |
|---|---|
| -0.89703060374286 | X36 X37 X38 |
| -0.28448203187155 | X36 X37 A38 B39 |
| -0.28448203187155 | X36 X37 B38 A39 |
| 0.08348704968869 | X36 A37 X38 B39 |
| 0.08348704968869 | X36 B37 X38 A39 |
| 0.07374053821242 | X37 X38 X39 |
| 0.06548869326992 | X36 X38 X39 |
| -0.05383556111760 | A36 X37 X38 B39 |
| -0.05383556111760 | B36 X37 X38 A39 |
| 0.03341790728550 | A36 X37 B38 X39 |
| 0.03341790728550 | B36 X37 A38 X39 |
| -0.02359527063297 | X36 A37 B38 X39 |

---

**eq-T S$_1$/S$_0$ MECI-3 (Open)**

---

Cartesian coordinates / Å

| | | | |
|---|---|---|---|
| C | 0.9095339727 | -0.0484331258 | -0.0145314402 |
| C | 2.4043536677 | -0.0444539669 | -0.3004398358 |
| C | -0.6140307484 | 1.2605330511 | -1.0733290568 |
| C | 0.1142566625 | -1.1501561702 | -0.3416282763 |
| C | 2.7170111627 | -0.1369487777 | -1.8004323764 |
| C | 3.0597019168 | 1.2034339509 | 0.2990071460 |
| C | -1.7322649527 | 1.0753231098 | -0.2360560362 |
| C | -2.1417125342 | -0.2268026002 | 0.0502701397 |
| C | -1.2831609853 | -1.3065251392 | -0.2228198041 |
| C | -3.5149561917 | -0.4695766576 | 0.6178650859 |
| H | 0.5862682481 | 0.5806494983 | 0.7942841497 |
| H | 2.8691181041 | -0.9107573943 | 0.1781022715 |
| H | -0.5162613603 | 0.6694718828 | -1.9631322573 |

```
H   -0.1166549753    2.2136853439   -1.1021302435
H    0.6281030632   -1.9970853961   -0.7783569712
H    2.3174378144    0.7237250566   -2.3298834921
H    3.7889243003   -0.1693223792   -1.9758097948
H    2.2879820292   -1.0284835047   -2.2474674697
H    2.6433039812    2.1096171065   -0.1342353788
H    4.1298929343    1.2109819630    0.1157456207
H    2.9071979689    1.2498611373    1.3730552513
H   -2.1539588949    1.9137546092    0.2885955467
H   -1.6910535089   -2.2986023907   -0.3061337975
H   -4.2145320376   -0.6366491780   -0.1976406605
H   -3.8650939060    0.3925009507    1.1739962064
H   -3.5360107075   -1.3399039953    1.2622727925
```

| | |
|---|---|
| $S_0$ energy / H | -387.82229947872599 |
| $S_0$ CI eigenvector | 0.58775178793951  X36 X37 A38 B39 |
| | 0.58775178793951  X36 X37 B38 A39 |
| | 0.53104751369019  X36 X37 X38 |
| | -0.09031680732821  X36 X38 X39 |
| | -0.07797737688523  X36 A37 B38 X39 |
| | -0.07797737688523  X36 B37 A38 X39 |
| | 0.05596380898023  X36 X37 X39 |
| | -0.04928821467268  X37 X38 X39 |
| | 0.02206405608562  X36 A37 X38 B39 |
| | 0.02206405608562  X36 B37 X38 A39 |
| $S_1$ energy / H | -387.82229908597662 |
| $S_1$ CI eigenvector | -0.82271966075544  X36 X37 X38 |
| | 0.38303160573690  X36 X37 A38 B39 |
| | 0.38303160573690  X36 X37 B38 A39 |

0.11641835123835  X36 X38 X39

0.07026407790796  X37 X38 X39

-0.06080567079592  X36 X37 X39

-0.04515824235417  X36 A37 X38 B39

-0.04515824235417  X36 B37 X38 A39

-0.03874787926089  X36 A37 B38 X39

-0.03874787926089  X36 B37 A38 X3

---

**eq-T cZc-EDOT Minimum**

---

| Cartesian coordinates / Å | 26 | | | |
|---|---|---|---|---|
| | C | 1.0962937617 | -0.7437084803 | 0.1187906338 |
| | C | 2.5784273855 | -0.9430739121 | -0.1005682561 |
| | C | -0.8491385437 | 2.0587091358 | -0.7375281197 |
| | C | 0.0999406563 | -1.2777773747 | -0.5793084452 |
| | C | 2.9316018990 | -1.9482684154 | -1.1973467000 |
| | C | 3.2538553380 | 0.4118305896 | -0.3650625402 |
| | C | -1.4801362514 | 1.3720170901 | 0.1984734119 |
| | C | -2.0269544013 | 0.0042657496 | 0.0506468334 |
| | C | -1.3343727055 | -1.0914261003 | -0.3493704851 |
| | C | -3.5024156684 | -0.0977276454 | 0.3684862025 |
| | H | 0.8541511774 | -0.0843802515 | 0.9359452259 |
| | H | 2.9846312340 | -1.3238424698 | 0.8368144877 |
| | H | -0.6298542426 | 1.6325425825 | -1.6998220723 |
| | H | -0.5221819891 | 3.0691135703 | -0.5659913635 |
| | H | 0.3339182122 | -1.9804019920 | -1.3627237568 |
| | H | 2.5764377184 | -1.6159904974 | -2.1683545036 |
| | H | 4.0086103012 | -2.0668852925 | -1.2634683732 |
| | H | 2.5030384298 | -2.9246786351 | -0.9971472383 |
| | H | 2.9066016755 | 0.8382236794 | -1.3016790344 |

|   |   |   |   |
|---|---|---|---|
| H | 4.3326280947 | 0.3018702704 | -0.4238711746 |
| H | 3.0349166896 | 1.1233115206 | 0.4253637654 |
| H | -1.7000717027 | 1.8659234599 | 1.1346346124 |
| H | -1.9162051507 | -1.9856578828 | -0.5103255262 |
| H | -3.7008369536 | 0.1864134829 | 1.4005579276 |
| H | -3.8705728381 | -1.1070557719 | 0.2280202815 |
| H | -4.0883221262 | 0.5691235901 | -0.2598557930 |

| | |
|---|---|
| $S_0$ energy / H | -387.92443007312096 |
| $S_0$ CI eigenvector | 0.98724719543351  X36 X37 X38 |
| | -0.09231252203294  X36 X37 X39 |
| | -0.07473813237284  X37 X38 X39 |
| | -0.05810477880187  A36 X37 B38 X39 |
| | -0.05810477880187  B36 X37 A38 X39 |
| | 0.03731851511057  X36 X37 A38 B39 |
| | 0.03731851511057  X36 X37 B38 A39 |
| | -0.03662211345440  X36 X38 X39 |
| $S_1$ energy / H | -387.73435296924430 |
| $S_1$ CI eigenvector | -0.68776779599788  X36 X37 A38 B39 |
| | -0.68776779599788  X36 X37 B38 A39 |
| | 0.12157726426136  A36 X37 X38 B39 |
| | 0.12157726426136  B36 X37 X38 A39 |
| | 0.09033090811472  X36 A37 B38 X39 |
| | 0.09033090811472  X36 B37 A38 X39 |
| | 0.05115726381170  X36 X37 X38 |
| | 0.04896846654400  X36 X37 X39 |
| | -0.03442345199243  X37 X38 X39 |
| | -0.02903790866638  X36 A37 X38 B39 |

−0.02903790866638  X36 B37 X38 A39

| | **eq-G+ S$_0$ Minimum (FC)** |
|---|---|
| Cartesian coordinates / Å | 26 |
| | C   0.6960482445  −0.0870689034   0.1032427960 |
| | C   2.1626863025  −0.0576844258  −0.3832934109 |
| | C  −0.0884986580   1.1743503570  −0.2878428753 |
| | C  −0.0793547894  −1.3184258211  −0.3060929493 |
| | C   3.0166369418  −1.1085996037   0.3367219916 |
| | C   2.3167036572  −0.1977258456  −1.9026816995 |
| | C  −1.5530235132   1.0607302408   0.0506945782 |
| | C  −2.1888588477  −0.1141951421   0.0128325073 |
| | C  −1.4302457422  −1.3135032371  −0.3048362342 |
| | C  −3.6681042849  −0.2510727812   0.2682471986 |
| | H   0.7336829485  −0.0940102295   1.1955738903 |
| | H   2.5576055164   0.9177123264  −0.1028197183 |
| | H   0.0016911040   1.3541740970  −1.3593550705 |
| | H   0.3463757605   2.0390329813   0.2046486454 |
| | H   0.4495607089  −2.2299324216  −0.5185427318 |
| | H   2.7126398569  −2.1199528478   0.0833499680 |
| | H   4.0620250710  −1.0068359534   0.0609856035 |
| | H   2.9468010749  −1.0026704737   1.4151857457 |
| | H   1.9456648592  −1.1559234419  −2.2535025973 |
| | H   3.3639117843  −0.1292760459  −2.1822904325 |
| | H   1.7855179837   0.5777640602  −2.4427714802 |
| | H  −2.0947893591   1.9657719398   0.2677627609 |
| | H  −1.9718511037  −2.2187961190  −0.5217577433 |
| | H  −4.1243692097   0.7083293507   0.4818543027 |
| | H  −3.8594285947  −0.9124520087   1.1092605924 |
| | H  −4.1710277119  −0.6787400518  −0.5955736373 |
| S$_0$ energy / H | -387.96128271615066 |
| S$_0$ CI eigenvector | -0.97490610199104  X36 X37 X38 |
| |  0.11644922646756  X36 X37 X39 |
| | -0.09740199834209  X36 X37 A38 B39 |
| | -0.09740199834209  X36 X37 B38 A39 |
| |  0.08649470651096  X36 X38 X39 |
| | -0.05806762497663  X36 A37 B38 X39 |
| | -0.05806762497663  X36 B37 A38 X39 |
| | -0.03149315218857  A36 X37 B38 X39 |
| | -0.03149315218857  B36 X37 A38 X39 |
| S$_1$ energy / H | -387.77296474415903 |
| S$_1$ CI eigenvector | -0.66612638392567  X36 X37 A38 B39 |
| | -0.66612638392567  X36 X37 B38 A39 |
| | -0.16489117153764  X36 A37 X38 B39 |
| | -0.16489117153764  X36 B37 X38 A39 |
| |  0.14235848364915  X36 X37 X39 |
| |  0.13452794941286  X36 X37 X38 |
| | -0.09922599578473  X36 X38 X39 |
| |  0.04801419686799  X36 A37 B38 X39 |
| |  0.04801419686799  X36 B37 A38 X39 |
| | -0.03528431505315  A36 X37 B38 X39 |
| | -0.03528431505315  B36 X37 A38 X39 |
| | -0.03316164040981  A36 X37 X38 B39 |
| | -0.03316164040981  B36 X37 X38 A39 |

**eq-G+ S$_1$/S$_0$ MECI-1 (Closed)**

| Cartesian coordinates / Å | C | 0.7583159085 | -0.1006263508 | 0.4949642215 |
|---|---|---|---|---|
| | C | 1.9487010256 | -0.1456607667 | -0.4810838568 |
| | C | 0.0362621951 | 1.2489130232 | 0.6182955523 |
| | C | -0.4000957131 | -0.9817307749 | 0.0920694465 |
| | C | 3.0843608934 | 0.7459294013 | 0.0315582008 |
| | C | 2.4451455340 | -1.5766496329 | -0.7138307309 |
| | C | -1.4294795874 | 0.9137438898 | 0.7936023980 |
| | C | -2.0935997272 | 0.1953150222 | -0.2301624815 |
| | C | -1.2901988639 | -0.6302770664 | -1.0040252554 |
| | C | -3.5988551832 | 0.1826037232 | -0.2949452518 |
| | H | 1.0962516366 | -0.4122297405 | 1.4806115447 |
| | H | 1.6004122949 | 0.2458075544 | -1.4326058947 |
| | H | 0.1762541409 | 1.8403927503 | -0.2865916253 |
| | H | 0.4127924833 | 1.8305480182 | 1.4510988390 |
| | H | -0.5440764187 | -1.9126056505 | 0.6231471687 |
| | H | 3.4748676744 | 0.3721392274 | 0.9752809843 |
| | H | 3.9062817777 | 0.7730530190 | -0.6774150585 |
| | H | 2.7549207141 | 1.7675084191 | 0.1912428782 |
| | H | 2.7521409499 | -2.0476113045 | 0.2173166375 |
| | H | 3.3003600918 | -1.5836978018 | -1.3825883053 |
| | H | 1.6777242467 | -2.1967451010 | -1.1682206727 |
| | H | -2.0081858679 | 1.4131702389 | 1.5502220112 |
| | H | -1.7213068391 | -1.3786644420 | -1.6428089561 |
| | H | -4.0031026839 | 1.1574858604 | -0.0419692561 |
| | H | -3.9948556205 | -0.5392439088 | 0.4136463770 |
| | H | -3.9349812458 | -0.1001352588 | -1.2851147592 |

| $S_0$ energy / H | -387.80940292764240 | | |
|---|---|---|---|
| $S_0$ CI eigenvector | -0.57871978062015 | X36 X37 A38 B39 | |
| | -0.57871978062015 | X36 X37 B38 A39 | |

|  |  | -0.53617385164606 X36 X37 X38 |
|  |  | 0.11386776958470 X36 A37 B38 X39 |
|  |  | 0.11386776958470 X36 B37 A38 X39 |
|  |  | -0.07835583232262 X36 A37 X38 B39 |
|  |  | -0.07835583232262 X36 B37 X38 A39 |
|  |  | 0.05916993786748 X36 X38 X39 |
|  |  | 0.02260994333456 X37 X38 X39 |

| S$_1$ energy / H | -387.80940272617028 |  |
|---|---|---|
| S$_1$ CI eigenvector |  | 0.81952993002138 X36 X37 X38 |
|  |  | -0.38119570043481 X36 X37 A38 B39 |
|  |  | -0.38119570043481 X36 X37 B38 A39 |
|  |  | -0.09724002381702 X36 X38 X39 |
|  |  | 0.08382713005827 X36 A37 X38 B39 |
|  |  | 0.08382713005827 X36 B37 X38 A39 |
|  |  | 0.07935555110863 X36 A37 B38 X39 |
|  |  | 0.07935555110863 X36 B37 A38 X39 |
|  |  | -0.03504285474656 X37 X38 X39 |

**eq-G+ S$_1$/S$_0$ MECI-2 (Closed)**

| Cartesian coordinates / Å | 26 |  |  |  |
|---|---|---|---|---|
|  | C | 0.8315692947 | 0.1936518889 | 0.2965853420 |
|  | C | 2.2488748655 | 0.0342300278 | -0.3042589839 |
|  | C | -0.0646043360 | 1.2102494815 | -0.4584503758 |
|  | C | 0.1274751178 | -1.1172893598 | 0.5143741097 |
|  | C | 3.2624147346 | -0.4860102298 | 0.7208815773 |
|  | C | 2.2727783115 | -0.8253841500 | -1.5744887382 |
|  | C | -1.5667438866 | 1.1714418026 | -0.0693820186 |
|  | C | -2.0160138756 | -0.1735338215 | -0.0062272574 |

|   |   |   |   |
|---|---|---|---|
| C | -1.2016932727 | -1.2840778700 | 0.4753470157 |
| C | -3.4129099395 | -0.5031336533 | -0.4366770891 |
| H | 0.9513282988 | 0.6005013115 | 1.3063549915 |
| H | 2.5607728611 | 1.0380565790 | -0.5840049119 |
| H | 0.0094683597 | 1.0448079098 | -1.5309233253 |
| H | 0.3310577267 | 2.2032964669 | -0.2803775304 |
| H | 0.7345817273 | -1.9740067991 | 0.7566192562 |
| H | 3.3340578954 | 0.1809041148 | 1.5743766335 |
| H | 2.9940964084 | -1.4710820752 | 1.0945018293 |
| H | 4.2516965260 | -0.5674029528 | 0.2809951803 |
| H | 2.0585066510 | -1.8682404251 | -1.3550400536 |
| H | 3.2539034249 | -0.7913808219 | -2.0374618744 |
| H | 1.5512536083 | -0.4870484672 | -2.3095571029 |
| H | -1.3983297640 | 1.0683958322 | 1.0628601462 |
| H | -1.6362366087 | -2.2542421784 | 0.6455213089 |
| H | -4.0928127451 | -0.0525943141 | 0.2794246599 |
| H | -3.6239923538 | -1.5647047145 | -0.5067538660 |
| H | -3.5893352794 | 0.0039729897 | -1.3780759565 |

| | |
|---|---|
| S$_0$ energy / H | -387.80693665837327 |
| S$_0$ CI eigenvector | -0.67990519954931  X36 X37 A38 B39 |
| | -0.67990519954931  X36 X37 B38 A39 |
| | -0.22899503431839  X36 X37 X38 |
| | -0.08242271308404  A36 X37 B38 X39 |
| | -0.08242271308404  B36 X37 A38 X39 |
| |  0.06073958417323  X36 A37 B38 X39 |
| |  0.06073958417323  X36 B37 A38 X39 |
| | -0.02504941402394  X36 A37 X38 B39 |
| | -0.02504941402394  X36 B37 X38 A39 |
| |  0.02176228949323  X37 X38 X39 |
| S$_1$ energy / H | -387.80693659693350 |
| S$_1$ CI eigenvector | 0.95481877791196  X36 X37 X38 |
| | -0.16296049640291  X36 X37 A38 B39 |
| | -0.16296049640291  X36 X37 B38 A39 |
| | 0.08705774046205  X36 A37 X38 B39 |
| | 0.08705774046205  X36 B37 X38 A39 |
| | -0.08112785137459  X37 X38 X39 |
| | -0.06804658048944  X36 X38 X39 |
| | -0.05979856312741  A36 X37 X38 B39 |
| | -0.05979856312741  B36 X37 X38 A39 |
| | -0.02002049944278  A36 X37 B38 X39 |
| | -0.02002049944278  B36 X37 A38 X39 |

## eq-G+ S$_1$/S$_0$ MECI-3 (Open)

Cartesian coordinates / Å

| | | | |
|---|---|---|---|
| C | 0.8393340240 | -0.1842733329 | 0.0702667985 |
| C | 2.3413296884 | -0.1757239922 | -0.1744777564 |
| C | -0.5134496424 | 1.2430574449 | -0.9860143405 |
| C | -0.0193656647 | -1.2093660117 | -0.3304106005 |
| C | 3.1105786672 | -0.1531246947 | 1.1549305307 |
| C | 2.8517825385 | -1.3127529898 | -1.0644544973 |
| C | -1.6916604615 | 1.1418784961 | -0.2118601578 |
| C | -2.2114439674 | -0.1278473010 | 0.0337061001 |
| C | -1.4289487822 | -1.2635637487 | -0.2356345117 |
| C | -3.6172771574 | -0.2808463004 | 0.5510067190 |
| H | 0.5242766161 | 0.4342377115 | 0.8904969234 |
| H | 2.5902854585 | 0.7546639658 | -0.6886958454 |
| H | -0.4382334611 | 0.6920281665 | -1.9045899679 |
| H | 0.0781506110 | 2.1413529522 | -0.9482154056 |
| H | 0.4299577159 | -2.0620520236 | -0.8159763421 |
| H | 2.9413193639 | -1.0671242670 | 1.7168988304 |
| H | 4.1797277991 | -0.0522195241 | 0.9870992124 |
| H | 2.7947026510 | 0.6807948467 | 1.7749008549 |
| H | 2.7103178607 | -2.2822464684 | -0.5947365760 |
| H | 3.9141255320 | -1.1935633428 | -1.2520552820 |
| H | 2.3489934244 | -1.3288369831 | -2.0268865176 |
| H | -2.0564048330 | 2.0022742172 | 0.3195550163 |
| H | -1.9062515518 | -2.2198941063 | -0.3581808740 |
| H | -4.2872514452 | -0.4717969091 | -0.2839040172 |
| H | -3.9576001172 | 0.6240130942 | 1.0403949788 |
| H | -3.7000937252 | -1.1123041585 | 1.2408739903 |

| | |
|---|---|
| $S_0$ energy / H | -387.82076950452392 |
| $S_0$ CI eigenvector | -0.68673998548323  X36 X37 A38 B39 |
| | -0.68673998548323  X36 X37 B38 A39 |
| |  0.19740638436937  X36 X37 X38 |
| | -0.08824148116916  X36 A37 B38 X39 |
| | -0.08824148116916  X36 B37 A38 X39 |
| $S_1$ energy / H | -387.82076928014169 |
| $S_1$ CI eigenvector |  0.95997457664482  X36 X37 X38 |
| | -0.14086373112353  X36 X38 X39 |
| |  0.13894945635338  X36 X37 A38 B39 |
| |  0.13894945635338  X36 X37 B38 A39 |
| | -0.08507873053137  X37 X38 X39 |
| |  0.07221049275525  X36 X37 X39 |
| | -0.05062706457381  X36 A37 X38 B39 |

-0.05062706457381  X36 B37 X38 A39

0.02550730321154  X36 A37 B38 X39

0.02550730321154  X36 B37 A38 X39

---

**eq-G+ cZc-EDOT Minimum**

| Cartesian coordinates / Å | 26 | | | |
|---|---|---|---|---|
| | C | 1.1069891144 | -0.7071434348 | 0.1435385351 |
| | C | 2.5967442049 | -0.8780873330 | -0.0446227292 |
| | C | -0.8185241000 | 2.1256866707 | -0.5383397457 |
| | C | 0.1373462659 | -1.1905767785 | -0.6249294365 |
| | C | 3.2136466248 | -1.5081232905 | 1.2136645063 |
| | C | 2.9961560918 | -1.6619931778 | -1.2953199812 |
| | C | -1.4972466021 | 1.3737901447 | 0.3107580425 |
| | C | -2.0211864692 | 0.0157608187 | 0.0450907957 |
| | C | -1.3049009896 | -1.0389914637 | -0.4182374975 |
| | C | -3.5030402028 | -0.1267083574 | 0.3145846343 |
| | H | 0.8335645302 | -0.1257240997 | 1.0085965066 |
| | H | 3.0151856688 | 0.1248179686 | -0.1377011611 |
| | H | -0.5372310951 | 1.7670350185 | -1.5118288134 |
| | H | -0.5130731037 | 3.1242137821 | -0.2797404200 |
| | H | 0.4008133456 | -1.8205237740 | -1.4592070886 |
| | H | 2.8462630536 | -2.5202705978 | 1.3556113861 |
| | H | 4.2961272098 | -1.5491988451 | 1.1367846848 |
| | H | 2.9652700445 | -0.9374212340 | 2.1033944019 |
| | H | 2.6298581219 | -2.6836014882 | -1.2549705960 |
| | H | 4.0774024363 | -1.7034303514 | -1.3817157185 |
| | H | 2.6087423078 | -1.2007468222 | -2.1976855500 |
| | H | -1.7781139413 | 1.8046241389 | 1.2615862602 |
| | H | -1.8710859596 | -1.9242776642 | -0.6627375201 |

|   |   |   |   |
|---|---|---|---|
| H | -3.7320318991 | 0.0763063957 | 1.3593564204 |
| H | -3.8538589751 | -1.1266271484 | 0.0886782979 |
| H | -4.0815956827 | 0.5785109230 | -0.2776982141 |

| | |
|---|---|
| S₀ energy / H | -387.92458684230434 |
| S₀ CI eigenvector | 0.98734756732215  X36 X37 X38 |
| | -0.09243563086125  X36 X37 X39 |
| | -0.07448575785208  X37 X38 X39 |
| | 0.05767436211655  A36 X37 B38 X39 |
| | 0.05767436211655  B36 X37 A38 X39 |
| | -0.03797202759569  X36 X38 X39 |
| | 0.03607817069294  X36 X37 A38 B39 |
| | 0.03607817069294  X36 X37 B38 A39 |
| S₁ energy / H | -387.73530515519013 |
| S₁ CI eigenvector | -0.68823877641986  X36 X37 A38 B39 |
| | -0.68823877641986  X36 X37 B38 A39 |
| | -0.11913239829799  A36 X37 X38 B39 |
| | -0.11913239829799  B36 X37 X38 A39 |
| | 0.09078221553536  X36 A37 B38 X39 |
| | 0.09078221553536  X36 B37 A38 X39 |
| | 0.04950738469998  X36 X37 X38 |
| | 0.04666626926509  X36 X37 X39 |
| | -0.03212382714577  X37 X38 X39 |
| | -0.03169745228226  X36 A37 X38 B39 |
| | -0.03169745228226  X36 B37 X38 A39 |

**Table S3.**

Coordinates, Energies and CI Eigenvectors for $S_0$ And $S_1$ at Critical Points Along the Ring-Closed and Ring-Open Pathways of αPH. For the CI eigenvectors, XYY indicates that the YYth molecular orbital is doubly occupied, and AYY/BYY indicate that the YYth molecular orbital is singly occupied with alpha or beta spin, respectively.

## Supplementary Movies

**All movies were generated from the coordinate expectation value of a single AIMS trajectory basis function propagating first on $S_1$ and then spawned on $S_0$ using $\alpha$(0.82)-SA2-CAS(6,4)-SCF/6-31G* and uPBE0/6-31G*, respectively.

1. (ax G+) aPH-CI1-cisL.mov
2. (ax T) aPH-CI1-cisL.mov
3. (eq G-) aPH-CI1-cisL.mov
4. (eq T) aPH-CI1-cisL.mov

These movies show the out-of-plane motion of $C_4$ in the ax/eq rotamers of aPH as the nuclear wavepacket propagates along the $S_1$ potential energy surface towards the CI-1 $S_1/S_0$ conical intersection (purple). The nuclear wavepacket moves away from the Frank-Condon region towards CI-1, where the $C_4$H bond pyramidalizes at approximately 90° to the molecular plane and the isopropyl group closely resembles its initial ax/eq configuration (see Fig. S8 for structural parameters). Upon relaxation to vibrationally "hot" aPH (grey), a carbon-carbon bond is formed between $C_7$ and $C_4$, resulting in a relatively stable cis-Ladderane species.

5. (ax G-) aPH-CI3-ZZDOT.mov
6. (ax G+) aPH-CI3-ZZDOT.mov
7. (ax T) aPH-CI3-ZZDOT.mov
8. (eq G-) aPH-CI3-ZEDOT.mov
9. (eq G+) aPH-CI3-ZEDOT.mov
10. (eq T) aPH-CI3-ZEDOT.mov

These movies show the elongation of $C_1$-$C_3$ bond in the ax/eq rotamers of aPH as the nuclear wavepacket propagates along the $S_1$ surface towards CI-3 $S_1$/$S_0$ conical intersection (purple). The nuclear wavepacket moves away from the Frank-Condon region towards CI-3, where $C_1$-$C_3$ increases to ~2.30 Å and the isopropyl group closely resembles its initial ax G- configuration (see Fig. S8 for relative values). Upon relaxation, a vibrationally "hot" mixture of ZZDOT/ZEDOT structures are observed on the $S_0$ ground electronic state (grey). The Woodward-Hoffman expected ZZDOT/ZEDOT photoproducts are clearly formed in each trajectory.

11. (ax G+) aPH-CI1-aPH.mov
12. (ax T) aPH-CI1-aPH.mov
13. (eq G-) aPH-CI1-aPH.mov
14. (eq T) aPH-CI1-aPH.mov

These movies show the out-of-plane motion of $C_4$ in the ax/eq rotamers of aPH as the nuclear wavepacket propagates along the $S_1$ potential energy surface towards the CI-1 $S_1$/$S_0$ conical intersection (purple). The nuclear wavepacket moves away from the Frank-Condon region towards CI-1, where the $C_4$H bond pyramidalizes at approximately 90° to the molecular plane and the isopropyl group closely resembles its initial ax/eq configuration (see Fig. S8 for structural parameters). Upon relaxation, vibrationally "hot" aPH is observed on the $S_0$ ground electronic state (grey) where axial-to-equatorial isomerization (and vice-versa) is observed in both conformers.

15. (ax G+) aPH-CI2-aPH.mov
16. (ax G-) aPH-CI2-aPH.mov
17. (ax T) aPH-CI2-aPH.mov
18. (eq G+) aPH-CI2-aPH.mov
19. (eq G-) aPH-CI2-aPH.mov
20. (eq T) aPH-CI2-aPH.mov

These movies show the out-of-plane motion of $C_7$ in the ax/eq rotamers of aPH as the nuclear wavepacket propagates along the $S_1$ potential energy surface towards the CI-2 $S_1$/$S_0$ conical intersection (purple). The nuclear wavepacket moves away from the Frank-Condon region towards CI-2, where the $C_7$H bond is oriented at approximately 90° to the molecular plane and the isopropyl group closely resembles its initial ax/eq configuration (see Fig. S8 for structural parameters). Upon relaxation, a vibrationally "hot" aPH is observed on the $S_0$ ground electronic state (grey).

21. (ax G+) aPH-CI3-aPH.mov
22. (ax G-) aPH-CI3-aPH.mov
23. (ax T) aPH-CI3-aPH.mov
24. (eq G+) aPH-CI3-aPH.mov
25. (eq G-) aPH-CI3-aPH.mov
26. (eq T) aPH-CI3-aPH.mov

These movies show the elongation of $C_1$-$C_3$ bond in the ax/eq rotamers of aPH as the nuclear wavepacket propagates along the $S_1$ surface towards CI-3 $S_1$/$S_0$ conical intersection (purple). The nuclear wavepacket moves away from the Frank-Condon region towards CI-3, where $C_1$-$C_3$ increases to ~2.30 Å and the isopropyl group closely resembles its initial ax G- configuration (see Fig. S6 for relative values). Upon relaxation, a vibrationally "hot" aPH is observed on the $S_0$ ground electronic state (grey).